%% file: DM-SUSY.tex
\documentclass[prd,twocolumn,preprintnumbers,amsmath,amssymb,nofootinbib]{revtex4}
\pdfoutput=1

\newcommand{\sla}[1]{#1\kern-0.55em/}
\usepackage[sort&compress]{natbib}
\usepackage{graphicx,epsfig}
\usepackage{dcolumn}
\usepackage{bm}
\usepackage{amsmath}
\usepackage{amssymb}
\usepackage{color}
\usepackage{float}
\usepackage{hyperref}

\definecolor{grey}{gray}{0.6}

\renewcommand\thefootnote{\fnsymbol{footnote}}

\def\MIN{{\sc Min}}
\def\MED{{\sc Med}}
\def\MAX{{\sc Max}}

\extrafloats{100}
\newcommand{\pbar}{\bar{p}}

\begin{document}

\hspace*{13.cm}\mbox{CERN-TH-2017-144}\\\vspace*{2.cm}

\title{Robustness of dark matter constraints and interplay\\ with collider searches for New Physics\vspace*{1.cm}} 

\author{A. Arbey$^{1,2,3,}$\footnote{Also Institut Universitaire de France, 103 boulevard Saint-Michel, 75005 Paris, France}}
\email{alexandre.arbey@ens-lyon.fr}

\author{M. Boudaud$^{4,}$}
\email{boudaud@lpthe.jussieu.fr} 

\author{F. Mahmoudi$^{1,2,3,*,}$}
\email{nazila@cern.ch}

\author{G. Robbins$^{1,2,}$}
\email{glenn.robbins@univ-lyon1.fr}

\affiliation{~\\$^1$ Univ Lyon, Univ Lyon 1, ENS de Lyon, CNRS, Centre de Recherche Astrophysique de Lyon UMR5574, F-69230 Saint-Genis-Laval, France\vspace*{0.2cm}\\
$^2$ Univ Lyon, Univ Lyon 1, CNRS/IN2P3, Institut de Physique Nucl\'eaire de Lyon UMR5822, F-69622 Villeurbanne, France\vspace*{0.2cm}\\
$^3$ CERN, Theoretical Physics Department, CH-1211 Geneva 23, Switzerland\vspace*{0.2cm}\\
$^4$ Laboratoire de Physique Th\'eorique et Hautes \'Energies (LPTHE), UMR 7589 CNRS \& UPMC, 4 Place Jussieu, F-75252 Paris, France\vspace*{1.cm}}

\begin{abstract}
We study the implications of dark matter searches, together with collider constraints, on the phenomenological MSSM with neutralino dark matter and focus on the consequences of the related uncertainties in some detail. 
We consider, inter alia, the latest results from AMS-02, Fermi-LAT and XENON1T.
In particular, we examine the impact of the choice of the dark matter halo profile, as well as the propagation model for cosmic rays, for dark matter indirect detection and show that the constraints on the MSSM differ by one to two orders of magnitude depending on the astrophysical hypotheses. On the other hand, our limited knowledge of the local relic density in the vicinity of the Earth and the velocity of Earth in the dark matter halo leads to a factor 3 in the exclusion limits obtained by direct detection experiments. We identified the astrophysical models leading to the most conservative and the most stringent constraints and for each case studied the complementarities with the latest LHC measurements and limits from Higgs, SUSY and monojet searches. 
We show that combining all data from dark matter searches and colliders, a large fraction of our supersymmetric sample could be probed. Whereas the direct detection constraints are rather robust under the astrophysical assumptions, the uncertainties related to indirect detection can have an important impact on the number of the excluded points.

\end{abstract}


\maketitle

\setcounter{footnote}{0}
\renewcommand\thefootnote{\arabic{footnote}}

\section*{}
\cleardoublepage

\section{Introduction}
\label{sec:intro}
\input{intro}

\section{Dark matter searches and uncertainties}
\label{sec:cosmo}

\input{framework_relic}

\subsection{Indirect detection}
\label{sec:indir}
\input{framework_indirect}

\subsection{Direct detection}
\label{sec:dir}
\input{framework_direct}

\section{Analyses}
\label{sec:analyses}
\input{analyses}

\section{Results}
\label{sec:results}
\input{results}

\section{Summary}
\label{sec:conclusion}
\input{conclusions}

\section*{Acknowledgements}
The authors are grateful to M. Battaglia for his advice for the treatment of the LHC results and for many useful discussions, and would like to thank P. Salati, J. Silk and J. Billard for useful discussions. MB acknowledges support from the European Research Council (ERC) under the EU Seventh Framework Program (FP7/2007-2013)/ERC Starting Grant (agreement n. 278234 -- NewDark project led by M. Cirelli).\vspace*{0.5cm}


\bibliographystyle{h-physrev5}
\bibliography{biblio}

\end{document}

%% file: intro.tex
The Large Hadron Collider (LHC) is currently operating at 13 TeV, with the main purpose of searching for new particles and new phenomena, in order to establish the first direct proofs of New Physics Beyond the Standard Model (BSM). The lack of a stable, neutral, massive and very weakly interacting dark matter particle (\textsc{WIMP}) in the Standard Model (SM) constitutes one of the main motivations for search for New Physics. The dark matter (DM) paradigm emerged in astrophysics and cosmology, through the observation of galaxy rotation curves and galaxy clusters, giving birth to the concept of dark matter haloes around galaxies. Since then, numerous observations have indicated that dark matter is probably cold, {\it i.e.} has small velocities. In addition, collisions of clusters such as the Bullet Cluster reveal that dark matter can be separated from the baryons \cite{Markevitch:2003at,Clowe:2003tk}. Furthermore, Cosmic Microwave Background (CMB) observations, and in particular those of the Planck Collaboration \cite{Ade:2015xua}, provide precise measurements of the cold dark matter density, which can be distinguished from the sub-dominant baryon and hot dark matter densities.

One of the most common hypotheses is that cold dark matter is made of new particles still undiscovered at colliders. Two types of experiments have been designed in order to discover dark matter particles. First, direct detection experiments are based on the assumption that dark matter particles interact weakly with matter, and in particular with nucleons. In view of the dark matter density in galactic haloes, it is assumed that a large number of such particles would cross the Earth at any time. Therefore, the design of direct detection experiments is basically to gather a large amount of crystals or gases in big tanks in order to measure the recoil energy of the nuclei when dark matter scatters with the nucleons. Second, dark matter indirect detection experiments aim at detecting the annihilation or decay products of dark matter particles. In particular, the density of dark matter should be larger in the galactic center making the annihilation of dark matter particles more probable. Dark matter annihilations can produce gamma rays, which can be observed for example with Cherenkov telescopes, or other SM particles which can populate the antiproton spectrum, that is observed in particle detectors. So far, none of the dark matter detection experiments have been able to find a solid evidence for dark matter.

In parallel, the LHC continues the quest for New Physics. In Summer 2012, the announcement of the discovery of the Higgs boson with a mass of 125 GeV during the 8 TeV run \cite{Aad:2012tfa,Chatrchyan:2012xdj} was a final step towards the completion of the Standard Model spectrum. Thereafter, the measurements of the properties of the Higgs boson have established its compatibility with the SM predictions. Further, no new particle has been discovered yet, in spite of the higher energy of 13 TeV available in run 2. Searches for supersymmetric or BSM particles, or exotic phenomena such as monojets, are ongoing, and strong limits on BSM models are obtained. Monojet searches are generally considered as dark matter searches at the LHC, since they aim at finding evidence for missing energy in the final states through the presence of an initial state radiated hard jet. This missing energy would correspond to invisible particles leaving no energy in the detectors, which may be the dark matter constituents.

Searches for supersymmetry (SUSY) are still the main focus of the ATLAS and CMS experiments. In particular, the minimal supersymmetric extension of the Standard Model (MSSM) constitutes an excellent playground to design new physics searches or study dark matter, with however a limitation for systematic studies due to the large number of free parameters. If $R$-parity is conserved, supersymmetric particles can only interact in pair with SM particles, so that the lightest supersymmetric particle (LSP) is stable and can constitute dark matter, provided it is neutral and weakly interacting. The lightest neutralino is generally considered to be an adequate dark matter candidate.

The phenomenological MSSM (pMSSM) \cite{Djouadi:1998di}, with its 19 parameters, is a good compromise  as it is the most general $R$-parity and CP-conserving MSSM scenario respecting minimal flavour violation, and has a manageable number of parameters to allow for systematic studies.
In this paper, we will study dark matter and collider constraints within the pMSSM with 19 independent parameters, assuming that the lightest neutralino ($\tilde{\chi}_1^0$, simply labelled $\chi$ in the following) is the LSP. This scenario is general enough so that our main conclusions can hold in other supersymmetric scenarios. 
There have been several studies combining the LHC limits and dark matter constraints in the pMSSM, which either aim at determining the excluded regions or perform global fits in order to find the preferred parameter regions (see e.g. \cite{Roszkowski:2014iqa,Cahill-Rowley:2014twa,Arbey:2015aca,Aad:2015baa,Bertone:2015tza,Athron:2017ard,Workgroup:2017lvb,Athron:2017yua} for some recent studies).
Instead, the focus here will be on the astrophysical and cosmological uncertainties that can affect the interpretation of the dark matter experiment results, and to show explicitly and quantify the impact of such uncertainties in the dark matter limits that are set on the MSSM parameters.

The rest of this paper is organised as follows. In Section~\ref{sec:cosmo}, we will review the theoretical framework of relic dark matter density, indirect and direct detections, and study the astrophysical and cosmological uncertainties that can affect them. In Section~\ref{sec:analyses}, we describe the methods used for our analysis of the pMSSM, for the collider constraints, as well as for direct and indirect detections, and evaluate the general consequences of the choice of astrophysical and cosmological assumptions. In Section~\ref{sec:results}, we show the results in the pMSSM, considering the dark matter observables and their uncertainties, collider constraints, and the complementarity between dark matter and collider results. We will also briefly discuss the prospects for dark matter experiments. Finally, the conclusions will be given in Section~\ref{sec:conclusion}.

%% file: framework_relic.tex
\subsection{Relic density}
\label{sec:relic}

The dark matter abundance has been measured in the framework of the standard cosmological model, and the Planck Collaboration has provided a precise evaluation of the cold dark matter density \cite{Ade:2015xua}:
\begin{equation}
 \Omega_c h^2 = 0.1188 \pm 0.0010 \;.
\end{equation}

Constraints on new physics scenarios which propose dark matter candidates can therefore be obtained by comparing the computed dark matter density to the Planck value. The standard assumption to compute the dark matter density is to consider that dark matter particles are thermal relics, {\it i.e.} were in thermal equilibrium in the early Universe and we observe today only the surviving part. A second assumption is that there is a single thermal relic candidate contributing to the dark matter density, which is generally the case in BSM scenarios where dark matter particles have to be stable, electrically neutral and very weakly interacting.

With these assumptions, the relic density can be obtained by considering that all the new physics particles were originally in thermal equilibrium. Then the expansion of the Universe, which lowers the temperature, eventually breaks this equilibrium, and the evolution of the number densities of all the new particles can be obtained using Boltzmann equations, in which the expansion of the Universe introduces a friction-like term and the collision terms include annihilations and co-annihilations of these new particles into SM particles. When the dark matter density is diluted enough so that the interactions become negligible, the relic density is frozen and becomes only diluted by the expansion of the Universe. A detailed description of the calculation can be found in \cite{Gondolo:1990dk,Edsjo:1997bg}.

Comparing the obtained relic density to the very precise dark matter measurement can lead to very strong constraints on new physics parameters.

In the MSSM with R-parity conservation, neutralino and gravitino constitute good dark matter candidates, provided they are the lightest supersymmetric particles. The gravitino is however produced non-thermally, and was considered recently in \cite{CahillRowley:2012cb,Arvey:2015nra,Ibe:2016gir}. We therefore focus on the case of the lightest neutralino. The co-annihilations have in this case a very important role. If the lightest neutralino is mainly bino, it interacts weakly and the annihilation cross section is small, leading to a large relic density. To obtain the Planck limit, it is necessary to have co-annihilations of the neutralino with SUSY particles which are close in mass in order to increase the effective (co-)annihilation cross section. Similarly, if the neutralino is mostly wino or Higgsino, it is accompanied by a chargino which is very close in mass, making the co-annihilations possible. Considering the Sommerfeld enhancement, a wino-like neutralino can have the correct relic density naturally for a mass of about 2.8 TeV in absence of other co-annihilations, and 1 TeV for a Higgsino \cite{Hisano:2006nn,Cirelli:2007xd}. Careful studies about the consequences of the Sommerfeld enhancement in the context of relic density and indirect detection in the MSSM can be found in \cite{Zavala:2009mi,Hryczuk:2010zi,Hryczuk:2011vi,Hryczuk:2014hpa,Beneke:2016ync,Beneke:2016jpw}.

Several assumptions can nevertheless limit the constraining power of the relic density constraint as will be discussed in the following.

\subsubsection{Higher order corrections}

The first uncertainties arise from the numerical calculations of the annihilation and co-annihilation cross-sections. Whereas in the simplest cases the calculation of the relic density relies on a few decay channels, in the most compressed scenarios of the MSSM, more than 3000 channels can get involved, severely limiting the calculation speed of relic density. For this reason, the cross-sections are generally considered at tree-level. Yet, in individual channels, higher-order corrections can lead to 30\% modification or more \cite{Baro:2007em}. However, in most cases, the relic density calculated at tree-level differs by less than 10\% from the one calculated at one-loop \cite{Baro:2009na,Harz:2012fz}. Therefore, in the general case, about 10\% uncertainty can be associated to tree-level calculations of the relic density.

\subsubsection{QCD equations of state}

A second limitation comes from QCD equations of state. Indeed, computing the relic density requires the knowledge of the number of effective degrees of freedom of radiation, which lead to energy and entropy content of the Universe. While it was originally thought that the primordial plasma could be treated as an ideal gas above the QCD phase transition temperature, non-perturbative studies showed that at high temperature, the ideal gas approximation does not work, and different models for this plasma have been studied \cite{Hindmarsh:2005ix,Laine:2006cp,Drees:2015exa}, leading to different sets of QCD equations of state. The consequences on the relic density are however rather mild and can modify it by a few percent.

\subsubsection{Early Universe properties}

In the usual calculation of relic density, the expansion of the Universe is considered to be dominated purely by the radiation density. This hypothesis can however be falsified in many extensions of the standard model of cosmology \cite{Kamionkowski:1990ni,Salati:2002md,Profumo:2003hq,Chung:2007cn,Arbey:2008kv}. Similarly, entropy injection or non-thermal production of dark matter particles can modify the relic density \cite{Moroi:1999zb,Giudice:2000ex,Fornengo:2002db,Gelmini:2006pq,Arbey:2009gt}. These modifications of the standard model of cosmology can result in a change of the relic density by orders of magnitude, but are more likely to increase it. As a consequence, the uncertainties due to these effects are completely dominating the relic density calculation over the previous uncertainties.

To be conservative, we add to the Planck measurement error a theoretical uncertainty of 10\%, and consider the 3.5$\sigma$ interval
\begin{equation}
 0.0772 < \Omega h^2 < 0.1604\;.
\end{equation}

Moreover, since a modification of the cosmological standard scenario can result in a large increase of the relic density, the lower dark matter density limit can be disregarded.

%% file: framework_indirect.tex
Dark matter particles hosted in galaxies are supposed to annihilate into SM particles to yield, after hadronisation and decay, nuclei, electrons, photons and neutrinos.
The emissivity of one of these particles $i$ injected by the annihilation of two DM particles is
\begin{equation}
Q(E_i, \textbf{x}) = \eta \left( \frac{\rho(\textbf{x})}{m_{\rm DM}} \right)^2 \sum_{i,j} \langle \sigma v \rangle B_j \frac{dN_i^j}{dE_i}\;,
\label{eq:Q_primary}
\end{equation}
where $\langle \sigma v \rangle$ is the thermal average annihilating cross section, $B_j$ the branching ratio of the annihilation channel $j$, ${dN_i^j}/{dE_i}$ the multiplicity of the particle $i$ and $\eta$ is equal to $1/2$ ($1/4$) for a Majorana (Dirac) type particle. The density distribution $\rho$ of dark matter particles is discussed in Sec.~\ref{sec:dist}.
Indirect detection experiments try to find an excess of those messengers on top of their astrophysical background. Even in absence of signals, these experiments provide useful information about the dark matter nature.

Antiparticle cosmic rays are regarded with great interest. Indeed, their astrophysical background is composed of secondary particles {\it i.e.} particles produced by the interaction of primary cosmic rays (mostly proton and helium nuclei) on the interstellar medium (mostly hydrogen and helium atoms). Hence, their background is feeble and relatively under control compared to other species. 
Antiprotons ($\bar{p}$) are the most abundant antinuclei in cosmic rays that could be produced by dark matter and their spectral shape is distinguishable from the astrophysical background. For a dark matter mass larger than a few GeV, the flux of antiprotons features a cut off at the dark matter mass.
The most accurate measurements of the $\bar{p}$ flux at the Earth was reported by the space-borne detectors PAMELA~\cite{Adriani:2012paa} and AMS-02~\cite{Aguilar:2016kjl}. The discovery of an excess around 100 GeV was recently claimed~\cite{Cuoco:2016eej,Cui:2016ppb}.

However, both secondary and primary antiprotons suffer from theoretical uncertainties which make the significance of such an excess uncertain.
On the one hand, the astrophysical background of secondary antiprotons is affected by the lack of knowledge of the antiproton production cross section from proton-proton and proton-helium interactions, leading to an uncertainty for the flux at the Earth of $\sim$ 50\% (see for example~\cite{diMauro:2014zea,Winkler:2017xor}).
On the other hand, the antiproton flux produced by DM is very sensitive to the DM profile, altering the primary antiproton flux at the Earth by up to a factor of 2--6~\cite{Cirelli:2013hv,Evoli:2011id}.
In addition, both secondary and primary antiprotons are sensitive to uncertainties related to their propagation throughout the Galaxy~\cite{Donato:2001ms,Donato:2003xg}. Astrophysical uncertainties on galactic properties as well as the production cross sections used for secondary cosmic rays are the main uncertainties for the determination of the propagation parameters.
The total uncertainty for the secondary antiproton component was assessed in~\cite{Giesen:2015ufa,Evoli:2015vaa} to be up to a factor 3 at $\sim 100$ GeV. Moreover, the total uncertainty for the DM signal was shown to be as large as a factor of about 20 in~\cite{Cirelli:2013hv} and 50 in~\cite{Evoli:2011id}.
These results show how the constraints on the DM particle annihilation cross sections are sensitive to astrophysical and nuclear uncertainties. In the following, we reconsider these uncertainties using the most recent cosmic ray propagation results as well as the most recent galactic mass models, and study their consequences within the MSSM.

Cosmic ray positrons could also be produced by the annihilation of DM particles. Above a few GeV, the astrophysical background of positrons is not under control as for antiprotons. As a matter of fact, the positron excess reported by AMS-02~\cite{2014PhRvL.113l1102A} could be explained by the presence of young and nearby pulsars. In addition, the lack of knowledge about these systems makes it difficult to distinguish this hypothesis from an exotic component to explain the data. Therefore, this channel is not much useful to derive constraints on dark matter properties when $m_{DM}$ is larger than a few GeV.

Compared to charged cosmic rays, gamma rays have the advantage of propagating straight ahead. This allows us to characterise the morphology of their sources and to observe regions where the dark matter particle density is expected to be large and to produce a sizeable flux. The Fermi-LAT space-borne telescope covers the GeV energy range whereas the ground based Cherenkov telescopes HESS~\cite{Abdallah:2016ygi}, MAGIC~\cite{ Aleksic:2013xea}, VERITAS~\cite{Aliu:2012ga} and HAWC~\cite{Albert:2017vtb} are sensitive to the TeV range.
Since the density of dark matter particle is peaked in the center of the galaxy, the galactic center is one of the best targets to look for a dark matter signal. Nevertheless, this region hosts important astrophysical activities and it is difficult to estimate both the astrophysical background and foreground. Indeed, the gamma ray excess exhibited in the Fermi-LAT data ~\cite{Hooper:2010mq,Hooper:2011ti,Daylan:2014rsa,Abazajian:2014fta,Lacroix:2014eea,Calore:2014nla} could be interpreted either by annihilating DM particles, or by the presence of millisecond pulsars or the remnants of the past activity of the supermassive black hole lying in the center of the galaxy~\cite{Lee:2014mza,Petrovic:2014uda}.
On the other hand, dwarf spheroidal galaxies are considered as very interesting targets to look for a dark matter signal. Indeed, these systems are expected to i) be dominated in mass by a DM component ii) exhibit feeble stellar activities and then a low astrophysical background. Despite the fact that the dark matter distribution and concentration inside these objects is still under debate, they provide one of the best bounds on the average annihilating cross section $\langle \sigma v \rangle$.

\subsubsection{Dark matter halo profiles}
\label{sec:dist}

Dark matter particles are assumed to be isotropically distributed in a spherical halo around the galactic center. 
The radial density profile of dark matter arising from cosmological simulations were parametrised by Navarro, Frenk and White (NFW)~\cite{Navarro:1995iw} as
\begin{equation}
\label{eq:NFW}
\rho_{\rm NFW}(r)=\rho_s \frac{r_s}{r}\left(1+\frac{r}{r_s}\right)^{-2}\;,
\end{equation}
where $r_s$ is the radius at which the logarithmic slope of the profile is $-2$ and $\rho_s$ the dark mater density normalisation.

The Einasto profile on the other hand is defined as
\begin{equation}
\rho_{\rm Ein}(r)=\rho_s \exp\left\lbrace -\frac{2}{\alpha} \left[\left(\frac{r}{r_s}\right)^{\alpha}-1\right] \right\rbrace
\end{equation}
and provides a better agreement with the latest simulations~\cite{Navarro:2008kc} and does not suffer from the central divergence of~(\ref{eq:NFW}).

The star activity occurring in the inner galaxy could sweep dark matter particles from the inner region, resulting in a core profile as observed in many galaxies. Such profiles were introduced by Burkert \textit{et al.}~\cite{Burkert:1995yz} with the parametrisation 
\begin{equation}
\rho_{\rm Bur}(r)=\frac{\rho_s}{\left(1+\frac{r}{r_s}\right)\left(1+{\left(\frac{r}{r_s}\right)}^2\right)}\;.
\end{equation}
The parameters $r_s$ and $\rho_s$ as well as the distance of the Solar system to the galactic center are determined by dynamical observations of the Galaxy. We have used the values determined by~\cite{2011MNRAS.414.2446M} for NFW, by~\cite{Catena:2009mf} for Einasto and by~\cite{Nesti:2013uwa} for Burkert as reported in Table~\ref{tab:profiles}.
%
\begin{table}[t!]
\begin{center}
\begin{tabular}{|c|c|c|c|c|}
\hline
Halo profile&$r_s$&$\rho_{s}$& $R_{\odot}$& $\rho_{\odot}$\\
&[kpc]&[GeV/cm$^3$]&[kpc]&[GeV/cm$^3$]\\
\hline
NFW&$19.6$&$0.32$&$8.21$&$0.383$\\
Einasto($\alpha=0.22$)&$16.07$&$0.11$&$8.25$&$0.386$\\
Burkert&$9.26$&$1.57$&$7.94$&$0.487$\\
\hline
\end{tabular}
\caption{Dark matter mass model parameters for NFW~\cite{2011MNRAS.414.2446M}, for Einasto~\cite{Catena:2009mf} and for Burkert~\cite{Nesti:2013uwa} profiles.}
\label{tab:profiles}
\end{center}
\end{table}
\vspace*{0.3cm}
%

\subsubsection{Cosmic ray propagation}
\label{sec:prop}

Following the work of~\cite{Boudaud:2014qra} (and reference therein), we describe the galaxy using the two-zone model. The first zone, in which the interstellar medium is homogeneously distributed, represents the galactic disc of half-height $h = 100$ pc. Atomic densities are taken to be $n_{\rm H} = 0.9$ cm$^{-3}$ and $n_{\rm He} = 0.1$ cm$^{-3}$. The disc is embedded inside a magnetic halo of half-height $L$ lying between 1 and 15 kpc. Both zones share the radius $R = 20$ kpc.
As cosmic rays travel across the galaxy, they are affected by many processes as a result of their interactions with the galactic magnetic field and the interstellar medium.
The scattering of cosmic rays on the galactic magnetic field is modelled by a homogeneous and isotropic diffusion in space. The diffusion coefficient reads $K(E) = \beta K_0 (\mathcal{R}/1 \, \rm{GV})$ where $\beta$ is the velocity of the particle and $\mathcal{R}$ the rigidity related to its momentum $p$ and its charge $q$ by $\mathcal{R} = p/q$.
Since the diffusion centers move with the Alfv\`{e}n waves velocity $V_a$, the second-order Fermi mechanism applies and cosmic rays undergo a diffusive reacceleration. This process can be modelled by a diffusion in energy space with coefficient $D(E) = (2/9) V_a^2 E^2 \beta^4 / K(E).$
Moreover, cosmic rays can interact with the interstellar medium, leading to energy losses (including ionisation and Coulomb interaction) and their destruction at rates $b$ and $\Gamma$, respectively.
Finally, cosmic rays undergo the effect of the galactic wind produced by supernova remnant explosions in the galactic disc. We assume the galactic wind to be homogeneous and perpendicular to the galactic disc, with velocity $\textbf{V}_c = \text{sign}(z) V_c \, \textbf{e}_z$. This process leads to the adiabatic cooling of cosmic rays, which enters as an additional term in the energy loss rate $b$.

Under a steady state and thin disc approximation, the density of cosmic rays per unit of space and energy $\psi \equiv dN/d^3 \! x  dE$ obeys the transport equation
\begin{equation}
\label{eq:full_transport}
\begin{aligned}
&\nabla \cdot \left[ \textbf{V}_c \, \psi(E, r, z) - K(E) \, \nabla \psi(E, r, z) \right]   +  \\ &\partial_E \left[ b(E,z) \, \psi(E, r, z) -  \, 2h \, \delta(z) \, D(E) \, \partial_{E}\psi(E, r, z) \right] + \\   &2h \, \delta(z) \, \Gamma \psi = Q(E, r, z)\;,
\end{aligned}
\end{equation}
where $Q$ represents the injection rate of cosmic rays in the galaxy.

The interstellar flux of cosmic rays at the Earth is given by $\Phi(E, \odot) = v/4\pi  \,  \psi (E, \odot)$. For more details on the resolution method of the transport equation, we refer the reader to~\cite{2001ApJ...555..585M,2004PhRvD..69f3501D}. In this way, a semi-analytical method was used in~\cite{2004PhRvD..69f3501D} to derive the benchmark \MIN, \MED, and \MAX\ propagation models presented in Table~\ref{tab:propa}. The \MED\ model corresponds to the best fit to the boron over carbon (B/C) ratio whereas the \MIN\ and \MAX\ sets of parameters define the lower and upper bounds for the primary $\bar{p}$ flux, consistent with the B/C ratio.
We emphasise that recent papers, based on synchrotron radio emission \cite{DiBernardo:2012zu,Bringmann:2011py,Orlando:2013ysa,Fornengo:2014mna}, on cosmic ray positrons \cite{DiMauro:2014iia} as well as on gamma rays \cite{Ackermann:2012pya}, find that the thin halo predicted by \MIN\ is disfavoured. 
Furthermore, in Ref.~\cite{Lavalle:2014kca} it was pointed out that secondary positrons can be used to improve the determination of the propagation parameters since they do not undergo exactly the same propagation processes as for nuclei.
Following this idea, in Ref.~\cite{Boudaud:2016jvj} the pinching method was used to compute properly the flux of secondary positrons below 10 GeV and derive stringent constraints on the propagation parameters, which rule out the models with $L < 4$ kpc at the $3 \sigma$ level, including the \MIN\ benchmark propagation model. As a result, the \MED\ model provides a conservative lower bound to the dark matter antiproton signal.
The recent B/C data reported by AMS-02 and their future studies would result in an improved determination of the parameters of the propagation models.

Finally, cosmic rays have to penetrate the heliosphere where they interact with the Solar wind and the Solar magnetic field. In this work, we use the forced-field approximation~\cite{1971JGR....76..221F} parametrised by the Fisk potential $\phi_{\rm F}$, to predict the flux of cosmic rays on the top of the atmosphere where they are measured by the space-borne detectors.

%
\begin{table}[t!]
\begin{center}
\begin{tabular}{|c||c|c|c|c|c|}
\hline
Model		& $\delta$			& $K_0$ [kpc$^2$/Myr]		& $L$ [kpc]		& $V_c$ [km/s]	& $V_a$ [km/s]\\
\hline 
\hline
\MIN  		& 0.85			& 0.0016					& 1				& 13.5			&  22.4 \\
\MED  		& 0.70			& 0.0112					& 4 	 		& 12	 		& 52.9 \\
\MAX  		& 0.46			& 0.0765					& 15 			& 5	 			& 117.6\\
\hline
\end{tabular}
\caption{Benchmark \MIN, \MED, and \MAX\ sets of propagation parameters \cite{2004PhRvD..69f3501D}.}
\label{tab:propa}
\end{center}
\end{table}
\vspace*{0.3cm}
%

%% file: framework_direct.tex
Direct dark matter searches aim at directly detecting \textsc{wimp}s via tiny energy deposits when they scatter off target atomic nuclei in ultra-sensitive, low background detectors. No convincing dark matter signal has been detected so far, however, limits on the \textsc{wimp}-nucleon cross-section are set by comparing the measured differential recoil rate per unit detector mass to the theoretical rate given by:
\begin{equation}
\frac{dR}{dE_{R}}=\frac{\rho_{0}}{m_{DM} m_{N}} \int^{}_{v>v_{min}} \mathrm{d}\mathbf{v} \, f(\mathbf{v})v \frac{d \sigma_{\chi -N}}{dE_{R}}\;,
\end{equation}
where $\rho_0$ is the local \textsc{wimp} density, $m_{DM}$ is the \textsc{wimp} mass, $M_N$ the target nucleus mass, and
$v_{min}=\sqrt{\frac{E_{th} m_{N}}{2 \mu^{2}}}$, with $E_{th}$ the recoil energy threshold and $\mu=\frac{m_{N} m_{DM}}{m_{N}+m_{DM}}$ the reduced mass of the \textsc{wimp}-nucleus system.
$f(\mathbf{v})$ is the \textsc{wimp} velocity distribution in the Earth's rest frame.\\

 Usually, the \textsc{wimp}-nucleus cross-section is decomposed in a spin-independent (SI) and a spin-dependent (SD) contributions in the zero momentum transfer limit:
\begin{equation}
\frac{d \sigma_{\chi-N}}{dE_R}=\frac{m_N}{2 \mu^{2} v^{2}} (\sigma_{\chi -N}^{SI}F^{2}_{SI}(E_R)+\sigma_{\chi -N}^{SD}F^{2}_{SD}(E_R))\;,
\end{equation}
where the $\mathbf{F}$ functions are form factors describing how the \textsc{wimp} interferes with the nucleon structure of the nucleus and depend on the recoil energy. The $\sigma_{\chi -N}$ are the \textsc{wimp}-nucleus cross sections at zero momentum transferred. The SI form factors are experimentally well known from the study of elastic electronic scattering on nuclei and are reasonably well approximated by Helm form factors \cite{Lewin:1995rx}, while the SD form factors are obtained from nuclear shell model calculations \cite{Klos:2013rwa}. Those functions can be subject to uncertainties at high recoil energy, but their study is beyond the scope of this paper.

The SI \textsc{wimp}-nucleus cross-section is given by
\begin{equation}
\sigma_{\chi -N}^{SI}=\frac{4 \mu^2}{\pi}{[Z f_{p} +(A-Z) f_{n}]}^{2}\;,
\end{equation}
where $Z$ and $(A-Z)$ are the number of protons and neutrons in the nucleus and $f_{p}$ and $f_{n}$ are the effective SI \textsc{wimp}-proton and \textsc{wimp}-neutron couplings. In the calculation of experimental cross section limits, the approximation $f_{p} \approx f_{n}$ is commonly used. Under this assumption, the \textsc{wimp}-proton and \textsc{wimp}-neutron SI cross sections are equal ($\sigma_{\chi -p}^{SI}\approx \sigma_{\chi -n}^{SI}$) and the limits then concern a general \textsc{wimp}-nucleon cross section $\sigma_{\chi -nucleon}^{SI}$. Since this cross section scales as $A^{2}$, heavy target nuclei, such as argon and xenon, are favoured.
In this context, the strongest limits on SI cross section, for $m_{DM}\gtrsim 10$ GeV, are given by xenon target experiments, the leader being currently the XENON1T experiment \cite{Aprile:2017iyp}.\footnote{The PandaX-II Collaboration more recently released new limits which are slightly more stringent than the XENON1T ones for WIMP masses larger than $\sim$100 GeV \cite{Cui:2017nnn}.} Argon target experiments, such as DarkSide-50 \cite{Agnes:2015ftt}, give limits which are two orders of magnitude weaker. 

The SD \textsc{\textsc{wimp}}-nucleus cross-section is given by
\begin{equation}
\sigma_{\chi -N}^{SD}=\frac{32 \mu^2 G_{F}^{2}}{\pi} \frac{J+1}{J}{[a_{p} \left<S_{p}\right> +a_{n} \left<S_{n}\right>}]^{2}\;,
\end{equation}
where $G_{F}$ is the Fermi constant, $J$ is the total spin of the nucleus, $f_{p}$ and $f_{n}$ are the effective SI \textsc{\textsc{wimp}}-proton and \textsc{\textsc{wimp}}-neutron couplings and $\left<S_{p,n}\right>$ are the average spin contributions from the protons and neutrons in the nucleus. On the one hand, to set constraints on the SD \textsc{\textsc{wimp}}-neutron cross section, nuclei with an even number of protons and an odd-number of neutrons are needed. In xenon target experiments, for instance, the spin is carried by neutrons in neutron-odd isotopes (${}^{129}Xe$, ${}^{131}Xe$). The best SD \textsc{\textsc{wimp}}-neutron cross section limit is currently given by the LUX experiment \cite{Akerib:2017kat}. On the other hand, to put constraints on the SD \textsc{\textsc{wimp}}-proton cross section, it is necessary to use nuclei with an odd number of protons. One of the strongest limits is given by the PICO-60 experiment, using $C_{3}F_{8}$ target \cite{Amole:2017dex}. This limit is in competition with the one coming from the 79-string IceCube detector \cite{Aartsen:2016zhm} that aims to detect a neutrino excess from the Sun. DM would be captured in the Sun through scattering on the hydrogen nuclei. The captured dark matter would then annihilate and produce neutrinos that would be detected by the IceCube experiment. Their limits on the cross section are calculated by assuming equilibrium between DM capture and its annihilation and depend strongly on the DM annihilation channel.

In the next few years, these limits are expected to be drastically lowered by experiments that will increase their total target mass and time of exposure, starting by XENONnT \cite{Aprile:2015uzo}, LZ \cite{Akerib:2015cja}, and reaching the neutrino background with DARWIN \cite{Aalbers:2016jon} in ten years or so.\\
Once the neutrino background is reached, if no dark matter particle is discovered by then, directional detection will be key to pursue the search for DM particles.

\subsubsection{Global and local dark matter densities}
\label{sec:local}
All the experimental limits are calculated using the benchmark value $\rho_{0}=0.3$ GeV/cm$^3$ for the local DM density, but recent studies give a best fit value closer to $0.4$ GeV/cm$^3$ \cite{Catena:2009mf, 2017MNRAS.465...76M, Nesti:2013uwa}. The uncertainties on the local density value are still quite large, one of the main source residing in the knowledge of the baryon density in the galaxy. There may also be a discrepancy between the value calculated from the study of the motion of nearby stars and the one calculated from a global fit of stellar dynamics over the galaxy, assuming a spherical dark matter halo. In our study, we will consider that the local DM density lies between $0.2$ and $0.6$ GeV/cm$^3$ (see \cite{Read:2014qva} for a complete review) and will choose three different values to test the impact of those uncertainties on the exclusions in our sample of points: $\rho_0=0.2$, $0.4$ and $0.6$ GeV/cm$^3$.

\subsubsection{Velocities}

Customarily, an isotropic Maxwellian distribution is assumed for the \textsc{WIMP} velocity distribution $f(\mathbf{v})$, with the galactic disk rotation velocity $v_{rot}$ being the most probable speed.
It corresponds to the Standard Halo Model describing the dark matter halo as a non-rotating isothermal sphere \cite{Freese:1987wu, Drukier:1986tm}. The canonical value for $v_{rot}$ is $220$ km/s but it is believed that it can range from 200 to 250 km/s \cite{Reid:2009nj,McMillan:2009yr,Bovy:2009dr}.\\ This velocity distribution is truncated at the escape velocity $v_{esc}$ at which a \textsc{\textsc{wimp}} can escape the galaxy potential well. Its value is subject to large uncertainties, $v_{esc}=500-600$ km/s, with a benchmark value $v_{esc}=544$ km/s \cite{Smith:2006ym}. However, for \textsc{\textsc{wimp}} masses $m_{DM} >10$ GeV, $v_{min}$ is relatively low. The velocity distribution is then integrated over a large range of velocities and $dR/dE_{R}$ is not sensitive to the tail of the distribution. Thus, the uncertainties on $v_{esc}$ should not impact our analysis.

Other halo models have been proposed, such as the King Model which describes the finite size of the halo and the gravitational interaction with ordinary matter in a more realistic way \cite{1966AJ.....71...64K, Chaudhury:2010hj} or such as triaxial halo models \cite{Evans:2000gr}. 
In this study, we will focus only on the uncertainties related to the Standard Halo Model, which is the most widely used in the literature.

%% file: analyses.tex
\subsection{MSSM Scans}
\label{sec:mssm}

We consider in this analysis the pMSSM, which is the most general $R$-parity and CP-conserving MSSM scenario with minimal flavour violation. 
It was shown in \cite{Chakraborty:2013si,Arbey:2014msa,Berger:2015eba} that CP violation does not have important consequences on the dark matter sector after imposing the experimental constraints from the electric dipole moments and the Higgs coupling measurements, so that the results presented in the following will remain valid also for CP violating scenarios.
We impose the lightest neutralino to be the lightest supersymmetric particle which constitutes dark matter, using the set-up presented in \cite{Arbey:2011un,Arbey:2011aa}. As the neutralino can be bino-like, wino-like, Higgsino-like or a mixed state, such a scenario allows for a large flexibility making our analysis of the astrophysical and cosmological uncertainties relatively general that can hold also in other dark matter models. We will not consider here the case of very light neutralinos that were studied in detail in \cite{Arbey:2012na,Belanger:2013pna,Boehm:2013gst,Arbey:2013aba}.

The pMSSM points are generated with SOFTSUSY \cite{Allanach:2001kg}, with a flat random sampling using the ranges given in Table~\ref{tab:pmssm} for the 19 parameters. After checking the theoretical validity of each point, we impose to have a neutralino LSP as well as a light Higgs of mass between 122 and 128 GeV. We then apply different constraints from the dark matter and collider experiments, which are described below. We do not intend to perform a robust statistical analysis as performed for example by the GAMBIT Collaboration~\cite{Workgroup:2017lvb,Workgroup:2017myk,Workgroup:2017htr}. Instead, the constraints are imposed separately at the $2 \sigma$ level, apart from for the Higgs sector where a likelihood analysis is used. As we do not aim at finding the preferred parameter regions, this choice will not affect the conclusions of our study.

\begin{table}[t!]
\begin{center}
\begin{tabular}{|c|c|}
\hline
~~~~Parameter~~~~ & ~~~~Range (in GeV)~~~~ \\
\hline\hline
$M_A$ & [50, 2000] \\
\hline
$M_1$ & [-3000, 3000] \\
\hline
$M_2$ & [-3000, 3000] \\
\hline
$M_3$ & [50, 3000] \\
\hline
$A_d=A_s=A_b$ & [-10000, 10000] \\
\hline
$A_u=A_c=A_t$ & [-10000, 10000] \\
\hline
$A_e=A_\mu=A_\tau$ & [-10000, 10000] \\
\hline
$\mu$ & [-3000, 3000] \\
\hline
$M_{\tilde{e}_L}=M_{\tilde{\mu}_L}$ & [0, 3000] \\
\hline
$M_{\tilde{e}_R}=M_{\tilde{\mu}_R}$ & [0, 3000] \\
\hline
$M_{\tilde{\tau}_L}$ & [0, 3000] \\
\hline
$M_{\tilde{\tau}_R}$ & [0, 3000] \\
\hline
$M_{\tilde{q}_{1L}}=M_{\tilde{q}_{2L}}$ & [0, 3000] \\
\hline
$M_{\tilde{q}_{3L}}$ & [0, 3000] \\
\hline
$M_{\tilde{u}_R}=M_{\tilde{c}_R}$ & [0, 3000] \\
\hline
$M_{\tilde{t}_R}$ & [0, 3000] \\
\hline
$M_{\tilde{d}_R}=M_{\tilde{s}_R}$ & [0, 3000] \\
\hline
$M_{\tilde{b}_R}$ & [0, 3000] \\
\hline
$\tan\beta$ & [1, 60]\\
\hline
\end{tabular}
\caption{pMSSM scan ranges.\label{tab:pmssm}}
\end{center}
\end{table}%

\subsection{Collider constraints}
\label{sec:collider}

Collider searches are very important to constrain the supersymmetric parameter space, but are also relevant for dark matter, through their correlations with the dark matter detection experiments. 

To the set of points in our analysis we apply constraints from LEP and Tevatron, from flavour physics, as well as the LHC constraints from the Higgs sector and supersymmetry and monojet direct searches. 

\subsubsection{LEP and Tevatron constraints}

LEP and Tevatron have provided very robust constraints on the mass of the supersymmetric particles \cite{Olive:2016xmw}. We apply to our set of points the limits summarised in Table~\ref{tab:LEP}.
\begin{table}
\centering
\begin{tabular}{|c|c|c|}
\hline
Particle & Limits & ~~~~~~Conditions~~~~~~\\
\hline\hline
$\tilde \chi^0_2$ & 62.4 & $\tan\beta < 40$\\
\hline
$\tilde \chi^0_3$ & 99.9 & $\tan\beta < 40$\\
\hline
$\tilde \chi^0_4$ & 116 & $\tan\beta < 40$\\
\hline
$\tilde \chi^\pm_1$ & 94 & $\tan\beta < 40$, $m_{\tilde \chi^\pm_1} - m_{\tilde \chi^0_1} > 5$ GeV\\
\hline
$\tilde{e}_R$ & 73 & \\
\hline
$\tilde{e}_L$ & 107 & \\
\hline
$\tilde{\tau}_1$ & 81.9 & $m_{\tilde{\tau}_1} - m_{\tilde \chi^0_1} > 15$ GeV\\
\hline
$\tilde{u}_R$ & 100 & $m_{\tilde{u}_R} - m_{\tilde \chi^0_1} > 10$ GeV\\
\hline
$\tilde{u}_L$ & 100 & $m_{\tilde{u}_L} - m_{\tilde \chi^0_1} > 10$ GeV\\
\hline
$\tilde{t}_1$ & 95.7 & $m_{\tilde{t}_1} - m_{\tilde \chi^0_1} > 10$ GeV\\
\hline
$\tilde{d}_R$ & 100 & $m_{\tilde{d}_R} - m_{\tilde \chi^0_1} > 10$ GeV\\
\hline
$\tilde{d}_L$ & 100 & $m_{\tilde{d}_L} - m_{\tilde \chi^0_1} > 10$ GeV\\
\hline
& 248 & $m_{\tilde \chi^0_1} < 70$ GeV, $m_{\tilde{b}_1} - m_{\tilde \chi^0_1} > 30$ GeV\\
& 220 & $m_{\tilde \chi^0_1} < 80$ GeV, $m_{\tilde{b}_1} - m_{\tilde \chi^0_1} > 30$ GeV\\
$\tilde{b}_1$ & 210 & $m_{\tilde \chi^0_1} < 100$ GeV, $m_{\tilde{b}_1} - m_{\tilde \chi^0_1} > 30$ GeV\\
& 200 & $m_{\tilde \chi^0_1} < 105$ GeV, $m_{\tilde{b}_1} - m_{\tilde \chi^0_1} > 30$ GeV\\
& 100 & $m_{\tilde{b}_1} - m_{\tilde \chi^0_1} > 5$ GeV\\
\hline
$\tilde{g}$ & 195 & \\
\hline
\end{tabular}
\caption{Constraints on the SUSY particle masses (in GeV) from searches at LEP and the Tevatron~\cite{Olive:2016xmw}.\label{tab:LEP}
}
\end{table}
One can note that LEP also provides limits for the lightest neutralino, but they can be evaded in specific cases and since our analysis is focussed on dark matter and the lightest neutralino, we prefer not to apply it. The neutralino mass will nevertheless be constrained by the light Higgs signal strength measurements, which can lead to stronger limits than LEP \cite{Djouadi:2005gj,Arbey:2012dq,Arbey:2012bp,Djouadi:2012rh,Chakraborti:2014gea,Chakraborti:2017dpu}.

\subsubsection{Flavour constraints}

Flavour constraints are complementary to the dark matter and collider searches. We consider here the three major decays, namely $B_s\to\mu^+\mu^-$, $B\to X_s \gamma$ and $B_u\to\tau\nu$ which capture the main constraints in the MSSM. $B_s\to\mu^+\mu^-$ is a rare transition which has a very strong constraining power. Indeed the scalar and pseudoscalar contributions lead to enhancements of its branching fraction proportional to $\tan^6\beta/M_A^4$, strongly constraining the large $\tan\beta$ and small $M_A$ parameter regions \cite{Huang:1998vb,Babu:1999hn,Ellis:2005sc,Eriksson:2008cx,Arbey:2012ax}. The inclusive decay $B\to X_s \gamma$ receives contributions from charged Higgs-top and chargino-stop loops, which also restrict the charged Higgs, stop and chargino masses in the large $\tan\beta$ regions. It is worth noting that the pseudoscalar and charged Higgs masses are connected at tree level by the relation
\begin{equation}
 M_{H^+}^2=M_A^2 + M_W^2\;,
\end{equation}
so that the pseudoscalar masses are also restricted. The third transition, $B_u\to\tau\nu$ is a tree-level leptonic decay which can be mediated by a $W$-boson or a charged Higgs. It also restricts the small $M_{H^+}$ and large $\tan\beta$ region. The value of the branching ratios of the three transitions is computed with SuperIso v3.7 \cite{Mahmoudi:2007vz,Mahmoudi:2008tp,Mahmoudi:2009zz}, and we apply the constraints shown in Table~\ref{tab:flavour}.

\begin{table}[t!]
\centering
\begin{tabular}{|c|c|c|}\hline
Observable & Experiment & SM prediction \\ \hline
$\mathrm{BR}(B_s \to \mu^+\mu^-) \times 10^9$ & $ 3.0 \pm 0.65$ \cite{Aaij:2017vad} & $3.54 \pm 0.27$\\ \hline
$\mathrm{BR}(B \to X_s \gamma) \times 10^4$ & $3.32 \pm 0.15$ \cite{Amhis:2016xyh} & $3.34 \pm 0.22$\\ \hline
$\mathrm{BR}(B_u \to \tau\nu_\tau) \times 10^4$ & $1.06\pm 0.19$ \cite{Amhis:2016xyh} & $0.82\pm 0.29$\\ \hline
\end{tabular}
\caption{Experimental results and the corresponding SM values for the flavour physics observables used in this work. The experimental data represents the most recent measurements or official combinations.\label{tab:flavour}}
\end{table}

\subsubsection{Higgs constraints}

Higgs searches provide strong constraints on the MSSM and the dark matter sector. On the one side, the measurements of the 125 GeV Higgs signal strengths provide constraints on the pMSSM Higgs sector parameters. Indeed, the Higgs mass is given by
\begin{eqnarray}
M_{h}^2 &\approx& M_Z^2 \cos^2 2{\beta} \left[1-\frac{M_Z^2}{{M_A}^2} \sin^2 2{\beta}\right] \\
&+& \dfrac{3 m_t^4}{2 \pi^2 v^2} \left[ \log \dfrac{{M_S}^2}{m_t^2} + \dfrac{{X_t}^2}{{M_S}^2}\left( 1 - \dfrac{{X_t}^2}{12{M_S}^2} \right)\right]\;,\nonumber
\end{eqnarray}
where only the leading terms are given. Here $M_S=\sqrt{M_{\tilde{t}_1}M_{\tilde{t}_2}}$ and $X_t=A_t-\mu\cot\beta$ are the most relevant parameters to achieve a mass of 125 GeV. 

The combined measurements of the Higgs mass by ATLAS and CMS from Run 1 gives \cite{Khachatryan:2016vau}
\begin{equation}
M_{h_\text{SM}} = 125.09 \pm 0.21 (\text{stat.}) \pm 0.11 (\text{syst.}) \;\text{GeV} \;. 
\end{equation}
However, the calculation of the Higgs mass in the MSSM is still subject to larger uncertainties (see for example \cite{Allanach:2004rh}), and we adopt the constraint:
\begin{equation}
 122\;\text{GeV} < M_{h_\text{SM}} < 128\;\text{GeV} \;. \label{higgs_mass}
\end{equation}

The measurement of the Higgs couplings is also particularly constraining. Indeed, the couplings of the Higgs bosons depend on the MSSM parameters as given at tree level in Table~\ref{tab:higgscoup}. $\alpha$ is the CP-even Higgs mixing angle:
\begin{equation}
 \alpha = \frac{1}{2}  \arctan \left( \tan(2 \beta) \, \frac{M_A^2 + M_Z^2 }{M_A^2 - M_Z^2 } \right) \;.
\end{equation}
In the decoupling limit, which corresponds to $M_A \gg M_Z$, the light Higgs couplings become SM-like.
\begin{table}[t!]
\begin{center}
\begin{tabular}{|c|c|c|c|}\hline
$\phi$ & $g_{\phi u\bar{u}}$ & $g_{\phi d\bar{d}}=g_{\phi \ell\bar{\ell}}$ & $g_{\Phi VV}$\\
\hline
$h$ & $\cos\alpha/\sin\beta {\color{grey}\,\to 1}$ & $-\sin\alpha/\cos\beta {\color{grey}\,\to 1}$ & $\sin(\beta-\alpha) {\color{grey}\,\to 1}$\\
\hline
$H$ & $\sin\alpha/\sin\beta {\color{grey}\,\to \cot\beta}$ & $\cos\alpha/\cos\beta {\color{grey}\,\to \tan\beta}$ & $\cos(\beta-\alpha) {\color{grey}\,\to 0}$\\
\hline
$A$ & $\cot\beta$ & $\tan\beta$ & 0\\
\hline
\end{tabular}
\end{center}
\caption{Tree level couplings of the Higgs bosons to quarks and vector bosons normalised to the SM couplings. The gray values correspond to the decoupling limit where $M_A \gg M_Z$.\label{tab:higgscoup}}
\end{table}
However, the couplings can receive higher-order corrections from the presence of supersymmetric particles, which can also lead to constraints on other MSSM parameters. LHC experiments have measured the signal strengths of different channels of the light Higgs boson, {\it i.e.} the product of the production cross sections times branching ratios. We use these measurements in our analyses, as given in Table~\ref{tab:Higgs}. The decays $h\to WW, ZZ, bb, \tau\tau$ provide direct constraints on the couplings given in Table~\ref{tab:higgscoup}. On the other hand, $h\to \gamma\gamma$ is a loop-level decay, in which the main contributions arise from top, stop, sbottom, chargino and charged Higgs loops~\cite{Haisch:2012re}. Its measurement is therefore particularly important to constrain the MSSM. Additionally, limits on the Higgs decays into supersymmetric particles can be set, which become particularly relevant if the dark matter particle is lighter than half of the Higgs mass. Since these decays to new particles participate to the total decay width of the light Higgs boson, they lower the branching ratios to SM particles.
\begin{table}[t!]
\centering
\begin{tabular}{|c|c|}
\hline
Channel & Experimental value\\
\hline\hline
$h\to\gamma\gamma$ & $1.14\pm0.19$\\
$h\to WW$ & $1.09\pm0.18$\\
$h\to ZZ$ & $1.29\pm0.26$\\
$h\to bb$ & $0.70\pm0.29$\\
$h\to \tau\tau$ & $1.11\pm0.24$\\
\hline
\end{tabular}
\caption{List of the Higgs signal strengths used in this analysis \cite{Khachatryan:2016vau}.}
\label{tab:Higgs}
\end{table}

The Higgs decay branching ratios and widths are computed using  {\tt HDECAY v6.51} \cite{Djouadi:1997yw}. The production cross sections are calculated using {\tt Sushi 1.5.0} \cite{Harlander:2012pb}, {\tt VV2H v1.10} and {\tt V2HV v1.10} \cite{Spira_codes}. The constraints are obtained through a likelihood analysis using the experimental and theoretical correlations from \cite{Khachatryan:2016vau} and \cite{Arbey:2016kqi}, respectively. Constraints are applied at the 95\% C.L.

\begin{table*}
\centering
\begin{tabular}{|c|c|c|c|}
\hline
Analysis & Target & 8 TeV & 13 TeV \\
\hline\hline
2-6 jets + MET & $\tilde{g}$, $\tilde{q}$ & 20 fb$^{-1}$ \cite{Aad:2015iea} & 13.3 fb$^{-1}$ \cite{ATLAS:2016kts}, 36.1 fb$^{-1}$ \cite{ATLAS:2017cjl} \\
\hline
7-11 jets +MET & $\tilde{g}$, $\tilde{q}$ & 20 fb$^{-1}$ \cite{Aad:2015iea} & 18.2 fb$^{-1}$ \cite{ATLAS:2016kbv}, 36.1 fb$^{-1}$ \cite{ATLAS:2017qzs}\\
\hline
2-6 jets + 1 lepton + MET & $\tilde{g}$, $\tilde{q}$ & 20 fb$^{-1}$ \cite{Aad:2015iea} & 14.8 fb$^{-1}$ \cite{ATLAS:2016lsr}\\
\hline
2, 3 leptons + MET & $\tilde{\chi}_2^0$, $\tilde{\chi}_1^\pm$, $\tilde{\ell}$ & 20 fb$^{-1}$ \cite{Aad:2015eda} & 13.3 fb$^{-1}$ \cite{ATLAS:2016uwq}, 36.1 fb$^{-1}$ \cite{ATLAS:2017uun}\\
\hline
jets + 0 lepton +MET & $\tilde{t}$ & 20 fb$^{-1}$ \cite{Aad:2015pfx} & 13.3 fb$^{-1}$ \cite{ATLAS:2016jaa}, 36.1 fb$^{-1}$, \cite{ATLAS:2017kyf}\\
\hline
jets + 1 lepton + MET & $\tilde{t}$ & 20 fb$^{-1}$ \cite{Aad:2015pfx} & 13.2 fb$^{-1}$ \cite{ATLAS:2016ljb}, 36.1 fb$^{-1}$ \cite{ATLAS:2017uun}\\
\hline
$b$-jets + 2 leptons + MET & $\tilde{t}$ & 20 fb$^{-1}$ \cite{Aad:2015pfx} & 13.3 fb$^{-1}$ \cite{ATLAS:2016xcm}, 36.1 fb$^{-1}$ \cite{ATLAS:2017tmd}\\
\hline
2 $b$-jets + MET & $\tilde{b}$, $\tilde{t}$ & 20 fb$^{-1}$ \cite{Aad:2015pfx} & 3.2 fb$^{-1}$ \cite{Aaboud:2016nwl}, 36.1 fb$^{-1}$, \cite{ATLAS:2017qih}\\
\hline
Monojet & MET & 20.3 fb$^{-1}$ \cite{Aad:2015zva} &  3.2 fb$^{-1}$ \cite{Aaboud:2016tnv}\\
\hline
mono-$Z,W$ & MET & 20.3 fb$^{-1}$ \cite{Aad:2013oja} & 3.2 fb$^{-1}$ \cite{TheATLAScollaboration:2015tsu}\\
\hline
\end{tabular}
\caption{List of ATLAS searches implemented in this analysis.}
\label{tab:LHC}
\end{table*}

Other relevant searches in the context of dark matter are searches for heavier Higgs bosons \cite{Djouadi:2005gj,Maiani:2012ij,Arbey:2013jla,Christensen:2012ei,Han:2013mga,Bechtle:2014ewa}. As can be seen from Table~\ref{tab:higgscoup}, in the limit when $M_A$ is large, the light Higgs couplings are SM-like, and compatible with the current data. The heavier states are therefore expected to be heavy. Nevertheless, the couplings of the $H/A$ to the $b$ quarks and $\tau$ leptons are enhanced by $\tan\beta$, so that it is possible to set strong limits in the small $M_A$ and large $\tan\beta$ region when searching for $(pp)bb\to H/A\to\tau\tau$. We use the results of CMS with 12.9 fb$^{-1}$ \cite{CMS:2016rjp}, and assess the exclusion by comparing the calculated cross section times branching ratio with the published tables. We note that it is sensitive to the same region which is probed by the branching ratio of $B_s\to\mu^+\mu^-$.

\subsubsection{LHC direct search constraints}

Direct searches from supersymmetric particles at the LHC provide amongst the most important constraints on the MSSM parameter space. We consider in our study the LHC searches presented in Table~\ref{tab:LHC}. Even if this list in not exhaustive, the most relevant searches for our study are considered, {\it i.e.} the channels with the highest sensitivity which are rather uncorrelated.

The SUSY direct searches correspond to final states with at least two SM particles and a large missing energy, carried by the invisible neutralinos. To assess the sensitivity of the LHC searches at 8 and 13 TeV, we generate inclusive samples of SUSY events with {\tt PYTHIA 8.150} \cite{Sjostrand:2006za,Sjostrand:2007gs}, using the CTEQ6L1 parton distribution functions \cite{Pumplin:2002vw}. {\tt Delphes 3.0} \cite{deFavereau:2013fsa} is then used to simulate the detector response and obtain the physics objects of the signal events. For each of the analyses, the signal selection cuts are applied to the simulated events, and the SM background events are taken from experimental publications. The CLs method~\cite{Read:2002hq} is used to obtain the 95\% confidence level (C.L.) exclusion in presence of background only.

Monojet and mono-$W,Z$ searches on the other hand have been designed in order to detect invisible particles in the final states through the detection of a hard jet emitted by the initial states. The basic idea is to search for a jet with high $p_T$ associated to a large missing $E_T$. The main background for monojet searches stems from $Z$ or $W$-boson and a jet, with the $Z$-boson decaying to neutrinos and the $W$-boson decaying to leptons which are missed by the detector. Considering models in which a single mediator relates the dark matter particles to the SM particles reveals that the LHC can have a competitive or even superior reach compared to the dark matter detection experiments \cite{Goodman:2010ku,Goodman:2010yf,Alves:2011wf,Abdallah:2015ter}. However, the simple description of dark matter production at the LHC based on a single mediator is not realistic with regard to concrete models such as the pMSSM, in which co-annihilations are favoured by the relic density constraints. Indeed, SUSY particles such as squarks or gluinos can be close in mass to the lightest neutralino, so that the production of two squarks or gluinos associated to a hard jet can still be seen as a monojet, because the jets produced in their decays would be soft enough to remain undetected \cite{Allanach:2010pp,Drees:2012dd,Cullen:2012eh,Arbey:2013iza,Arbey:2015hca}. In addition, several mediators can be involved. As a consequence, the single mediator limits cannot be recast in the pMSSM in a simple way. 

To study the exclusion by the monojet and mono-$W,Z$ searches, we use MadGraph 5 \cite{Alwall:2014hca} to compute the full $2\to3$ matrix elements for all the combinations of $pp \to \tilde{q}/\tilde{g} + \tilde{q}/\tilde{g} + j/W/Z$, $pp \to \tilde{\ell} + \tilde{\ell} + j/W/Z$ and $pp \to \tilde{\chi} + \tilde{\chi} + j/W/Z$, where $\tilde{q}$ refers to a squark of any type and generation, $\tilde{g}$ to the gluino, $\tilde{\ell}$ to any type of sleptons, $\tilde{\chi}$ to any electroweakino. $j$ corresponds to a hard jet as required for the monojet searches, and $W/Z$ for mono-$W,Z$ searches. As for the SUSY searches, we adopt the CTEQ6L1 parton distribution functions, hadronisation is performed using PYTHIA 8.150, and detector simulation with DELPHES 3.0. The cuts, selection efficiencies, acceptances and backgrounds for the 8 and 13~TeV runs are taken from the experimental publications cited in Table~\ref{tab:LHC}. In addition, as the systematic uncertainties can have an important effect on these limits \cite{Dreiner:2012sh,Arbey:2013iza,Baer:2014cua}, we account for them by adding a 30\% uncertainty to the cross sections.

\subsection{Dark matter constraints}

\subsubsection{Indirect detection}

The annihilation cross sections necessary for the interpretation of indirect detection data are calculated with MicrOMEGAs \cite{Belanger:2001fz,Belanger:2006is,Belanger:2013oya}, and PPPC4DMID \cite{Cirelli:2010xx} is used for the antiproton and gamma spectra.

\begin{figure*}
\begin{center}
\includegraphics[width=0.45\textwidth]{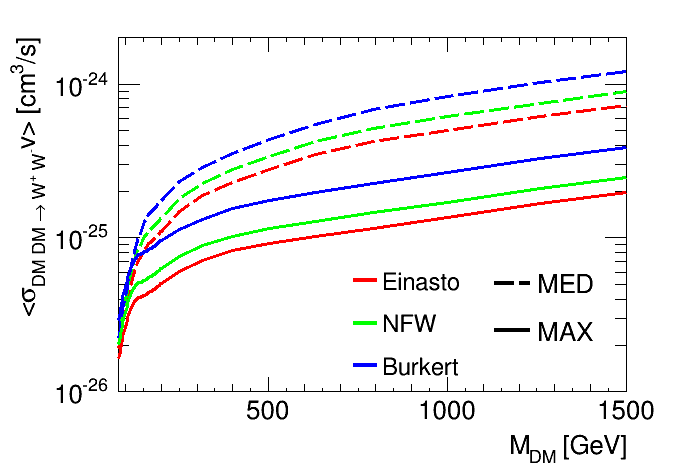}\includegraphics[width=0.45\textwidth]{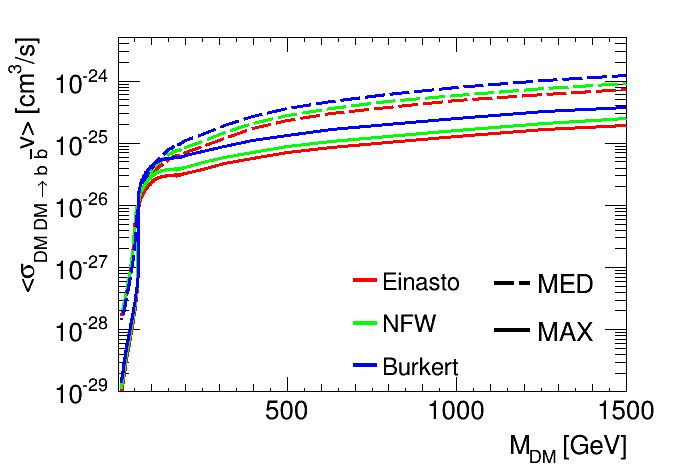}\\
\includegraphics[width=0.45\textwidth]{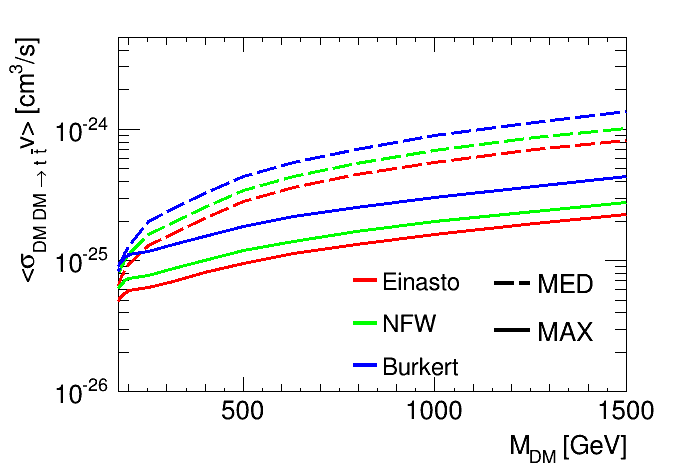}\includegraphics[width=0.45\textwidth]{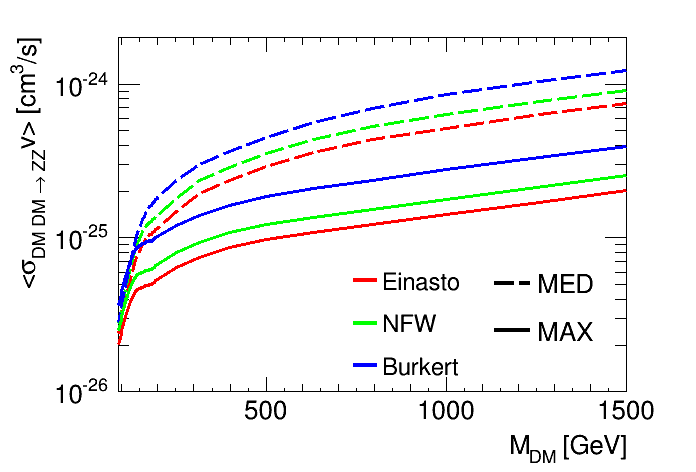} 
\caption{95\% C.L. upper limit of the neutralino annihilation cross section into $W^{+}W^{-}$ (upper left), $b \bar{b}$ (upper right), $t\bar{t}$ (lower left) and $ZZ$ (lower right), derived from AMS-02 antiproton data.}
\label{fig:pbar_limits}
\end{center}
\end{figure*}

\paragraph{Antiprotons}
We derived constraints on the dark matter annihilation cross section $\langle \sigma v \rangle$ from the cosmic ray antiproton flux measured by PAMELA~\cite{Adriani:2012paa} as well as AMS-02~\cite{Aguilar:2016kjl}, following the procedure described in~\cite{Boudaud:2014qra}.
Secondary antiprotons constitute the astrophysical background. They are mostly produced by the interaction of primary proton and helium cosmic ray nuclei on the hydrogen and helium atoms lying in the interstellar medium, and their production rate is given by
\begin{equation}
\begin{aligned}
& Q_{\pbar}^{\rm II}(\mathbf{x} , T_{\pbar})  =  4 \pi  \, (1 + N_{\rm IS}) {\sum_{i = {\rm p, He}}}  \; {\sum_{j = {\rm H, He}}}  \\ & \times {\displaystyle \int_{T^{0}_{i}}^{+ \infty}} dT_{i} \,
{\displaystyle \frac{d\sigma_{ij \to {\pbar} X}}{dT_{\pbar}}}(T_{i} \! \to \! T_{\pbar}) \, n_j (\mathbf{x}) \, \Phi_{i}(\mathbf{x} , T_{i}) \;,
\label{eq:Q_secondary}
\end{aligned}
\end{equation}
where $T_i$ is the kinetic energy of the nucleon $i$. The differential cross section $d\sigma_{ij \to {\pbar} X}/{dT_{\pbar}}$ is computed from the one for proton-proton interactions taken from~\cite{diMauro:2014zea} and the threshold $T_p^0$ of this reaction is taken to be $7m_p$. The factor $N_{\rm IS}$ takes into account the production of antineutrons decaying into antiprotons. Note that the bulk of the antiprotons are produced in proton-proton reactions.
Using the retropropagation technique, we computed the fluxes $\Phi_i(\mathbf{x})$ everywhere in the Galaxy from the fluxes $\Phi_i(\odot)$ measured at the Earth by PAMELA~\cite{Adriani2011} or AMS-02~\cite{Aguilar2015c_bis,Aguilar2015b_bis} when deriving constraints from the PAMELA or AMS-02 antiprotons flux, respectively.
We further renormalise the production rate (\ref{eq:Q_secondary}) by the factor $A$ which takes into account the energy dependent uncertainties on the production cross section as well as on the antineutron yield based on the analysis of~\cite{diMauro:2014zea}.

The production rate $Q_{\pbar}^{\rm I}$  of primary antiprotons produced by the annihilation of two dark matter particles into the channel $j$ is given by the expression (\ref{eq:Q_primary}) where the energy distribution of antiprotons per annihilation $dN_{\pbar}^j/dT_{\pbar}$ is taken from PPPC4DMID.
The propagation of primary and secondary antiprotons is computed with the semi-analytical scheme described in Sec.~\ref{sec:prop} for the \MED\ and \MAX\ sets of propagation parameters. The inelastic but non-annihilating interactions of antiprotons with the interstellar medium (tertiary component) are treated as in~\cite{Boudaud:2014qra}.
Practically, we apply the same procedure described in~\cite{Boudaud:2014qra} to derive the 95\% C.L. upper limit on the annihilation cross section $\langle \sigma v \rangle$, considering the secondary antiproton production uncertainty factor $A$ and the Fisk potential $\phi_{\rm F}$ as profiling parameters.

The 95\% C.L. upper limits on the annihilation cross section derived from the AMS-02 data are plotted in Fig.~\ref{fig:pbar_limits} with respect to the dark matter mass for the $W$, $b$, $t$ and $Z$ annihilation channels. These limits were computed for the \MED\ (dashed) and \MAX\ (solid) propagation models and for the Einasto (red), NFW (green) and Burkert (blue) galactic mass models.
The primary antiproton flux is maximised (minimised) by the \MAX\ (\MED) model and the upper bound value of $\langle \sigma v \rangle$ is thus minimised (maximised).
In addition, the DM density is much larger in the Galactic center for the cuspy Einasto and NFW profiles than for the Burkert one. As a result, the annihilation rate integrated out over the Galaxy is higher for the former, leading to more stringent constraints (even if the local DM density is larger for the latter).
As expected, the limits derived using the \MED\ model characterised by a smaller value of the halo size $L$, is less sensitive to the shape of the halo profile since $L$ is the ``propagation horizon".
In any case, the theoretical uncertainties coming from the poor knowledge of the propagation parameters is larger (up to a factor 4 on $\langle \sigma v \rangle$) than the one arising from the choice of the DM profile (up to a factor 2).
We found that for all annihilation channels and for the whole energy range we consider, the upper limit of $\langle \sigma v \rangle$ is maximised for the Burkert-\MED\ couple and minimised for the Einasto-\MAX\ one, providing an assessment of the theoretical uncertainties of our limits.
Note that the constraints for the $b$ quark channel become very stringent when $m_{DM}$ falls below ~50 GeV, excluding the thermal relic cross section up to 3 orders of magnitude at 10 GeV.

\begin{figure}[t!]
\begin{center}
\includegraphics[width=\columnwidth]{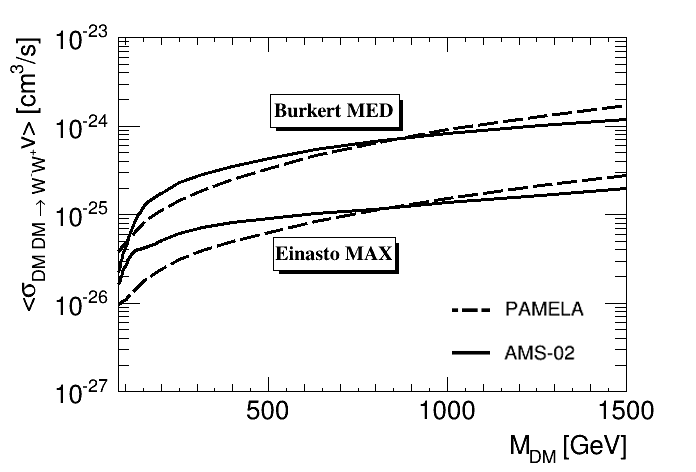}
\caption{Compared DM annihilation cross section upper limit from PAMELA (dashed) and AMS-02 (solid) antiproton data for the $W$ boson channel with Burkert-\MED\ and Einasto-\MAX.}
\label{fig:pbar_limits_ams_pamela}
\end{center}
\end{figure}

For the sake of consistency, we performed the same analysis using the PAMELA proton, helium and antiproton data. The comparison between the results for the $W$ boson channel obtained with AMS-02 and PAMELA data are plotted in Fig.~\ref{fig:pbar_limits_ams_pamela} for the Burkert-\MED\ and Einasto-\MAX\ cases.
For $m_{DM} \lesssim 1$ TeV, the constraints derived from PAMELA data are more stringent than the AMS-02 ones. This can be understood by the fact that below $\sim$1 TeV, the proton flux measured by PAMELA is larger than the one reported by AMS-02 by a factor of up to 10\%, leading to a larger yield of secondary antiprotons and thus a smaller room left for the primary component. The proton fluxes reported by the two experiments become similar above $\sim$1 TeV and in this regime, the experimental errors of AMS are much smaller than the PAMELA ones, leading to slightly stronger constraints for the more recent experiment.
In the following of this paper, we consider only the results obtained using the AMS-02 data since they are more recent and they provide globally more conservative results. 

As pointed out in~\cite{Giesen:2015ufa} and in this analysis, one of the main uncertainties on the constraints derived from the cosmic ray antiprotons arise from the lack of knowledge of the propagation parameters. Hopefully, this uncertainty will substantially shrink since the AMS-02 Collaboration have recently released the long-awaited boron to carbon ratio~\cite{PhysRevLett.117.231102}, crucial measurement for the determination of the galactic transport properties.

\paragraph{Gamma rays}
We now turn to a combined analysis of the 19 confirmed dwarf spheroidal galaxies (dSphs) recently observed by Fermi-LAT \cite{Fermi-LAT:2016uux}.

For each dSphs and for each model point, we calculated the expected gamma ray flux per energy bin from dark matter annihilation
\begin{align}
\Phi(\Delta \Omega, E_{min}, E_{max})=& \sum _{\substack{ \text{annihilation}\\ \text{channels}}} \frac{1}{4 \pi} \frac{<\sigma v>_{\text{channel}}}{2 m_{DM}^{2}} \nonumber\\
& \times  \int_{E_{min}}^{E_{max}} \left(\frac{dN_{\gamma}}{d E_{\gamma}}\right)_{\text{channel}}d E_{\gamma} \nonumber\\
 &\times\underbrace{ \int_{\Delta \Omega} \int_{l.o.s} \rho_{DM}^{2}(r(l)) dl d \Omega}_{J-factor}\;,
\end{align}
where $\left(\frac{dN_{\gamma}}{d E_{\gamma}}\right)_{\text{channel}}$ is the gamma ray spectrum produced from dark matter annihilation, which depends on the dark matter mass and its annihilation channel and is obtained by interpolating the spectra tabulated in the PPPC4DMID \cite{Cirelli:2010xx,Ciafaloni:2010ti}. The energy bins are those indicated in \cite{Fermi-LAT:2016uux}.\\
We were able to compute a delta-log likelihood for each of the points using the tabulated bin-by-bin likelihoods released by the Fermi-LAT Collaboration for each target \cite{GLASTurl} and we excluded points at the 95\% C.L. We include statistical uncertainties on the J-factors of each dwarf spheroidal galaxy by adding an additional J-factor likelihood term, as prescribed by the Fermi-LAT Collaboration in their study. Those J-factors were calculated assuming a NFW profile, but previous work showed that the limits calculated with other halo profiles differed only by $\sim$30\%, the strongest difference being for Burkert halo profile \cite{Ackermann:2015zua}. One of the largest uncertainties on these limits seems to reside in the choice of the dSphs sample used in the analysis. As pointed out in \cite{Fermi-LAT:2016uux}, adding galaxies with low-significance excesses, such as Reticulum II and Tucana III, can weaken significantly these limits. Assessing the effects of such uncertainties seems to be very delicate and we will use these limits only as comparison with the constraints coming from antiprotons.\\

In addition, we considered the limits given by the HESS Collaboration \cite{Abdallah:2016ygi}. As they do not use the same set of parameter values for the DM halo profiles as ours, we renormalised their limits following the J-Factors calculated for our different halo profiles NFW, Einasto and Burkert to be consistent with the rest of our study. The strongest limit being obtained for the NFW profile, we noticed that it barely reaches the distribution of our points without excluding any.

\subsubsection{Direct detection}

We calculated the \textsc{WIMP}-nucleon effective couplings $f_{p}$ and $f_{n}$ with MicrOMEGAs \cite{Belanger:2008sj,Belanger:2013oya}. In our sample of points, the approximation $f_{p} \approx f_{n}$ used in the calculation of experimental SI cross section limits is reasonable for Higgsino-like neutralinos, but is not necessarily correct for other neutralino types. There is a simple way to cope with this problem. The experimental \textsc{WIMP}-nucleon limits can be described more generally in terms of \textsc{WIMP}-nucleus limits, averaged over all the target isotopes, and renormalised to a \textsc{WIMP}-nucleon limit in the case $f_{p}=f_{n}$. Consequently, for a given point, the appropriate quantity to be compared to the experimental limit is:
\begin{equation}
\sigma_{\chi -nucleon}^{SI}(A)= \sigma_{\chi -p}^{SI} \frac{\sum_{i} \eta_{i} \mu_{A_{i}}^{2}[Z+(A_{i}-Z) f_{n}/f_{p}]^{2}}{\sum_{i} \eta_{i} \mu_{A_{i}}^{2}A_{i}^{2}}
\end{equation}
where the subscript $i$ stands for the various isotopes present in the experiment and $\eta_{i}$ is their corresponding abundance. These quantities depend on the target nucleus and are, a priori, different for xenon and argon. However, in our sample of points, we noticed that the relative difference $\delta=\left|\frac{\sigma_{\chi -nucleon}^{SI}(Xe)-\sigma_{\chi -nucleon}^{SI}(Ar)}{\sigma_{\chi -nucleon}^{SI}(Xe)}\right|$ was quite small, verifying $\delta \lesssim 10$\% ($\delta \lesssim 1$\% for the great majority of the points). The limits coming from xenon and argon experiments can then be easily compared, the XENON1T limit being the strongest one for our points.

Concerning the SD cross section limits, such problems do not exist. For the \textsc{WIMP}-neutron cross section, we will apply the limit given by the LUX experiment on our sample of points and for the \textsc{WIMP}-neutron cross section, we will use the one given by the PICO-60 experiment. We also tested the limits given by IceCube, using the $W^{+}W^{-}$ channel which is dominant for our wino-like and Higgsino-like neutralinos, and verified that the points excluded by the IceCube limit were already excluded by XENON1T or PICO-60.

\begin{figure}[t!]
\includegraphics[width=\columnwidth]{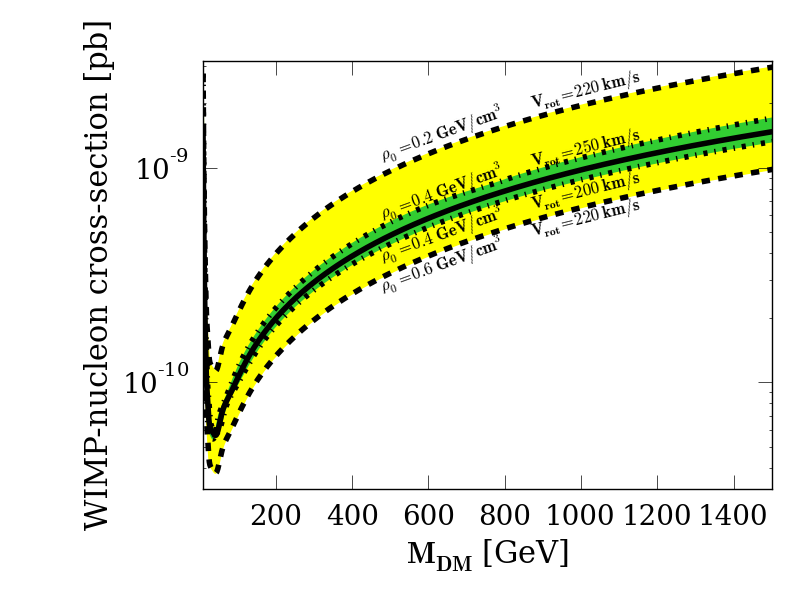}
\caption{XENON1T 90\% C.L. spin-independent \textsc{WIMP}-nucleon cross section upper limit for $\rho_{0}=0.4$ GeV/cm$^{3}$ and $v_{rot}=220$ km/s (black plain line). Uncertainties on these values are shown by varying independently the DM local density (yellow band) and the disc rotation velocity (green band).}
\label{dd_uncertainties}
\end{figure}

In addition, we examine how the uncertainties on the local dark matter density and on the disc rotation velocity impact these limits. By rescaling the cross section coordinates, we obtain the limits for the three local density values $0.2$, $0.4$ and $0.6$ GeV/cm$^{3}$. In order to test the impact of $v_{rot}$ uncertainties, we proceeded to a variable substitution in the integral of the velocity distribution appearing in the calculation of the differential recoil rate per unit detector mass. To perform such a calculation, it was necessary to consider that $v_{esc}\approx \infty$. This approximation induces errors only for low \textsc{WIMP} masses that are not concerned by our study. We were then able to rescale the upper limits originally calculated with $v_{rot}=220$ km/s for two other values $v_{rot}=200$ and $250$ km/s. Basically, taking smaller values for $v_{rot}$ shifts the limit to the right relative to $m_{DM}$, and taking larger values shifts it to the left. The impact of $\rho_{0}$ and $v_{rot}$ uncertainties on the XENON1T 90\% C.L. upper limit is shown in Fig.~\ref{dd_uncertainties}. The uncertainties on $v_{rot}$ within the considered values have a small impact compared to the local density uncertainties. Moreover, the uncertainties on $v_{rot}$ have a mild influence on the neutralino type of the excluded points and change the fraction of excluded points by less than $1$\%. For these reasons, and for sake of simplicity, we keep, in the rest of this study, the benchmark value $v_{rot}=220$ km/s and vary only the local dark matter density value.

%% file: results.tex
We consider only model-points which have the lightest neutralino as LSP and dark matter particle, as described in section~\ref{sec:mssm}. More than 20 million pMSSM points have been generated randomly with parameters in the ranges given in Table~\ref{tab:pmssm}. Prior to studying the effects of the different constraints, we impose the light Higgs mass to comply with the interval given in Eq.~(\ref{higgs_mass}).

\begin{figure}[t!]
  \centering
  \includegraphics[width=0.6\linewidth]{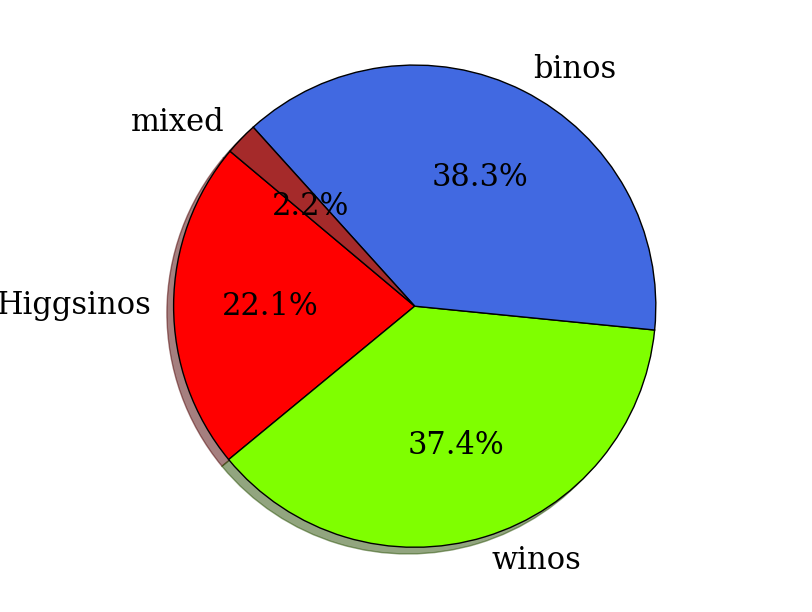}
  \caption{Fractions of neutralino 1 types in our scan after imposing only the light Higgs mass constraint.\label{pie:initial}}
\end{figure}

In the following, the neutralino 1 (denoted $\chi$) will be said to be bino-/wino-/Higgsino-like if it is composed of 90\% of bino-/wino-/Higgsino component, respectively, or mixed state otherwise. In Fig.~\ref{pie:initial}, the composition of our sample of pMSSM points after imposing the light Higgs mass interval is shown. Bino-like $\chi$ are the most represented points in our sample, followed by the winos and Higgsinos, with an almost equal share of each component. The fraction of mixed states is negligible.

\subsection{Relic density constraints}
\label{sec:mssm-relic}

\begin{figure}[t!]
\begin{center}
\includegraphics[width=\columnwidth]{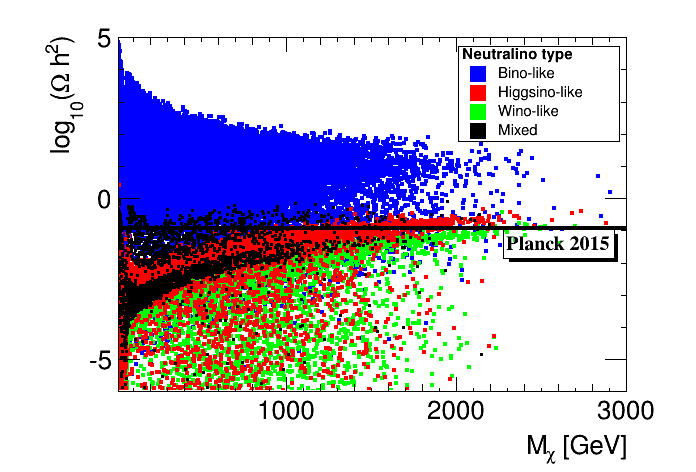} 
\caption{Neutralino relic density as a function of the neutralino 1 mass, for the different neutralino types. The central value of the Planck dark matter density is shown for comparison.\label{fig:relic_types}}
\end{center}
\end{figure}

We first consider the relic density constraint. The value of the neutralino relic density is computed with SuperIso Relic v3.4 \cite{Arbey:2009gu,Arbey:2011zz}. In Fig.~\ref{fig:relic_types}, the relic density is shown as a function of the neutralino 1 mass, for the different types. Bino-like neutralinos 1 have in general  large relic densities, above the Planck measurement. This can be explained by the smaller couplings of the binos with SM particles, which leads to smaller annihilation cross sections and therefore larger relic densities. On the other hand, the Higgsino-like $\chi$ give smaller relic densities which are close to the Planck measurements for $\chi$ masses around 1.3 TeV. The wino-like $\chi$ tend to have even smaller relic densities, and the Planck line is naturally reached for a mass of 2.7 TeV. The line at about 90 GeV in the figure corresponds to cross section enhancements through a $Z$-boson resonance, which lower the relic density.

\begin{figure}[t!]
\begin{center}
\includegraphics[width=\columnwidth]{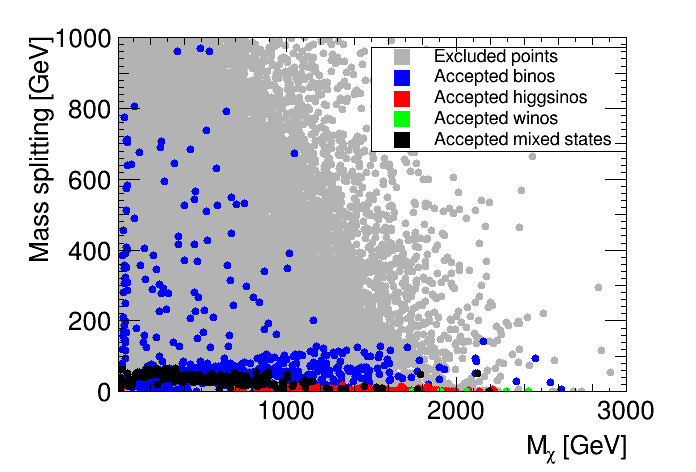} 
\caption{Points respecting both sides of the Planck 2015 relic dark matter density measurement in the mass splitting between the neutralino and the next lightest supersymmetric particle and the neutralino mass parameter plane.\label{fig:relic_masssplitt}}
\end{center}
\end{figure}

Imposing both the upper and lower relic density bounds generally leads to a selection of scenarios with co-annihilations, for which the mass splitting of the neutralino 1 with the next-to-lightest supersymmetric particle is small, or of scenarios where $\chi$ annihilations are enhanced through a resonance of the $Z$-boson or one of the neutral Higgs bosons. This is demonstrated in Fig.~\ref{fig:relic_masssplitt}. The valid points require in general small mass splitting, apart from some spread binos with larger mass splittings, which have a heavy Higgs boson or $Z$-boson resonance. For the case of winos, the small mass splitting is due to a chargino with a mass very close to the $\chi$ mass. For the Higgsino case, both the chargino 1 and the neutralino 2 have masses close to the neutralino 1 mass.

As discussed in Section~\ref{sec:relic}, we consider only the upper bound of the Planck dark matter density interval, which favours light wino- and Higgsino-like $\chi$, and bino-like $\chi$ with strong co-annihilations.

\subsection{Indirect detection constraints}
\label{sec:mssm-id}

\subsubsection{Constraints from AMS-02 and Fermi-LAT}

We consider the constraints from AMS-02 antiproton and Fermi-LAT gamma ray data, which probe specific dark matter annihilation channels. For both sets of constraints, the most important parameters are the $\chi$ annihilation cross sections into specific channels. Annihilations to $WW$ and $b\bar{b}$ are particularly interesting in the context of the pMSSM.

\begin{figure}[t!]
\begin{center}
\includegraphics[width=\columnwidth]{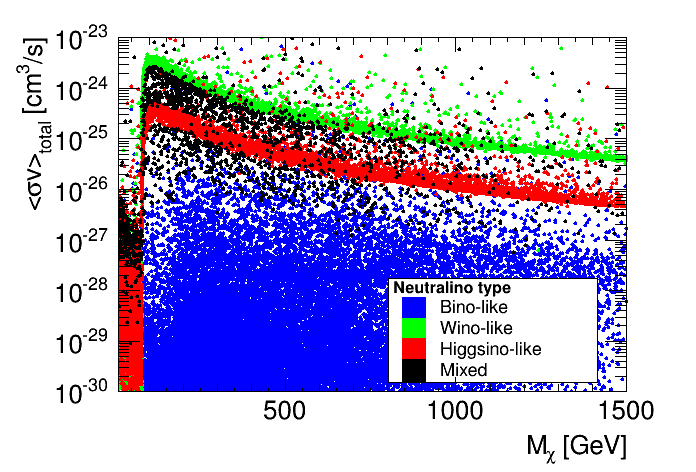} 
\caption{Total annihilation cross section as a function of the neutralino 1 mass for the different neutralino types.\label{fig:id_types}}
\end{center}
\end{figure}

In Fig.~\ref{fig:id_types}, the total annihilation cross section times velocity $\langle\sigma v\rangle_{\rm tot}$ is shown as a function of the neutralino 1 mass, for the different $\chi$ types.  $\langle\sigma v\rangle_{\rm tot}$ is the sum of all the $\sigma v$ of the different channels. The wino- and Higgsino-like neutralino 1 regions form two separate strips. The different types of neutralinos 1 have specific main decay channels: binos annihilate mainly into $t\bar{t}$, $b\bar{b}$, and in a lesser extent into $Wh$, $Zh$ and $\tau\tau$, Higgsinos into $WW$ and $ZZ$, and winos into $WW$, when the decay channels are open. When the above-mentioned channels are closed because of a small neutralino 1 mass, the $\chi$ mostly decays to $b\bar{b}$ and $\tau\tau$, and less frequently into $c\bar{c}$ and $s\bar{s}$, independently from their type. As seen earlier, winos more strongly annihilate than the other $\chi$ types, followed by the Higgsinos. The binos, apart from the case of a resonant annihilation, are more weakly annihilating and are far below the experimental limits.

\begin{figure}[t!]
\begin{center}
\includegraphics[width=\columnwidth]{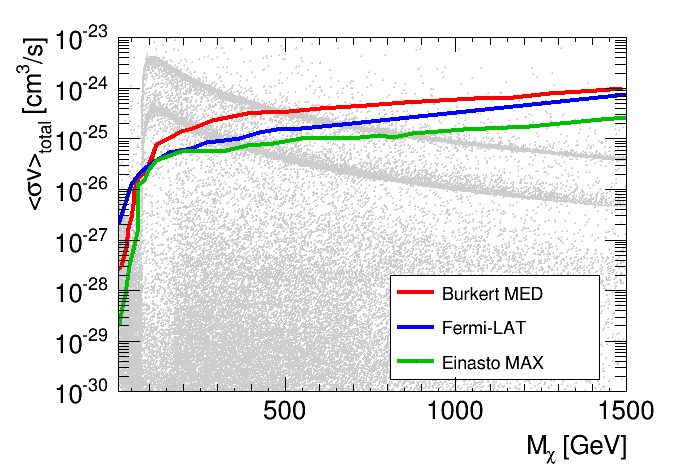} 
\caption{Points excluded by Fermi-LAT gamma ray and AMS-02 antiproton data in the total annihilation cross section vs. neutralino 1 mass parameter plane. The points above the red line are excluded by AMS-02 data in the conservative case with Burkert profile and \MED~propagation model, above the blue line by the Fermi-LAT data, and above the green line by AMS-02 data in the stringent case with Einasto profile and \MAX~propagation model.\label{fig:id_ams}}
\end{center}
\end{figure}

In Fig.~\ref{fig:id_ams}, the exclusion by Fermi-LAT and AMS-02 is shown in the $\langle\sigma v\rangle_{\rm tot}$ vs. neutralino 1 mass parameter plane. In order to quantify the uncertainties related to indirect detection, we consider separately the most conservative limits, {\it i.e.} obtained using Burkert profile and \MED~propagation model, and the most stringent ones, {\it i.e.} using Einasto profile and \MAX~propagation model. The conservative limits lead to the exclusion of neutralinos 1 with masses between 90 and 550 GeV, which are mainly wino-like. The stringent limits exclude points with $\chi$ masses between 0 and 850 GeV. In the small mass region, as well as for masses above 90 GeV, the stringent exclusion limit is strengthened by one order of magnitude in comparison to the conservative case. The stringent case excludes large zones of the wino strip, and of the Higgsino one in a lesser extent. AMS-02 alone brings very strong constraints in the stringent case, beyond the Fermi-LAT limits.

\subsubsection{Connections with relic density}

Indirect detection constraints may be considered to be redundant with the relic density constraint. This is generally true for simplified dark matter models \cite{Abdallah:2015ter}, because the relic density is directly related to the annihilation cross sections. However, in a complete model such as the MSSM, the value of the relic density is often led by the co-annihilations, especially when both the upper and lower bounds of the Planck dark matter density measurements are applied. This was already shown in Fig.~\ref{fig:relic_masssplitt}.

\begin{figure}[t!]
\begin{center}
\includegraphics[width=\columnwidth]{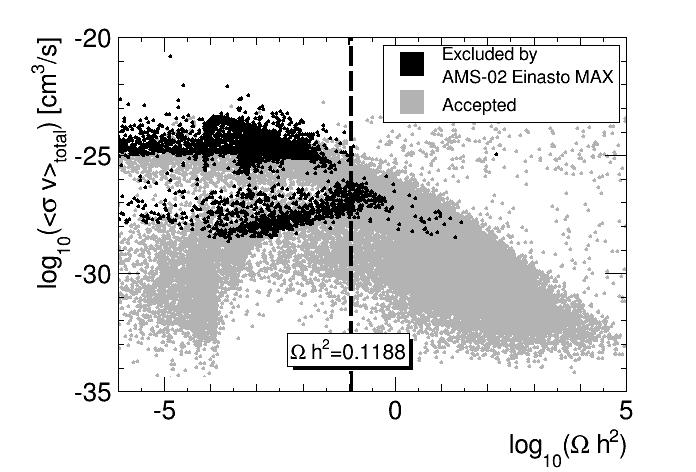} 
\caption{Total annihilation cross section as a function of the relic density. The vertical dashed line corresponds to the central value of the Planck dark matter density.\label{fig:relic_ID}}
\end{center}
\end{figure}

Yet, there is a strong complementarity between indirect detection and relic density, as shown in Fig.~\ref{fig:relic_ID}. Considering the gray points, we see an anti-correlated region where the relic density increases when the annihilation cross section decreases, which is due to the relation between the relic density and the annihilation cross sections. This region is largely excluded by the upper Planck bound. The points with small relic density have in general efficient co-annihilations, which reduces the relic density. While these points are far from being excluded by the Planck upper bound, they can be probed by the stringent AMS-02 limits obtained using the Einasto profile and \MAX~propagation model. This clearly shows the complementarity between indirect detection and relic density constraints.

\begin{figure}[t!]
\includegraphics[width=\columnwidth]{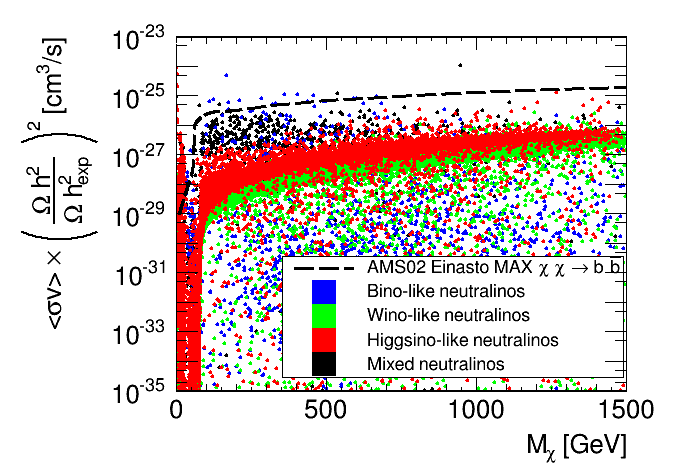}
\caption{Points excluded by Fermi-LAT gamma ray and AMS-02 antiproton data in the total annihilation cross section vs. neutralino 1 mass parameter plane, where the total annihilation cross section is rescaled by the relic density. The AMS-02 upper limit with Einasto profile and \MAX~propagation model for the $b\bar{b}$ channel is plotted for comparison.\label{fig:ID_rescaled}}
\end{figure}

In Section~\ref{sec:relic}, we discussed how the relic density constraint can be falsified. One of the possibilities is that the dark matter density measured by Planck is made only in part of neutralinos, the rest being made of other types of particles or more exotic objects. In such a case, galactic haloes would also be composed of different types of dark matters. Assuming that the mixture of dark matters is in the same proportion in galaxies as in the large scale Universe, the neutralino relic density is smaller than the measured dark matter density, and the dark matter density in galactic haloes has to be rescaled by the ratio of the neutralino relic density over the dark matter density, hence impacting the indirect detection limits. This is done in Fig.~\ref{fig:ID_rescaled}, in the total annihilation cross section vs. neutralino 1 mass parameter plane, where the total annihilation cross section is rescaled by the neutralino relic density over the measured dark matter density. Such a rescaling strongly weakens the indirect detection limits. Indeed, even using the most stringent AMS-02 constraints, only a very few points in the low mass region are still excluded, mostly in the $b\bar{b}$ channel. The large negative impact of the rescaling is due to the fact that the constraints from indirect detection scale as the squared density, leading to a strong loss of sensitivity.

\subsection{Direct detection constraints}
\label{sec:mssm-dd}

\subsubsection{Constraints from XENON1T, LUX and PICO-60}

Contrary to relic density and indirect detection, which mainly depend on the annihilation and co-annihilation cross sections, direct detection relies on the scattering cross section of neutralino 1 with nucleons. Direct detection is therefore complementary to indirect detection and relic density.

\begin{figure}[t!]
\begin{center}
\includegraphics[width=\columnwidth]{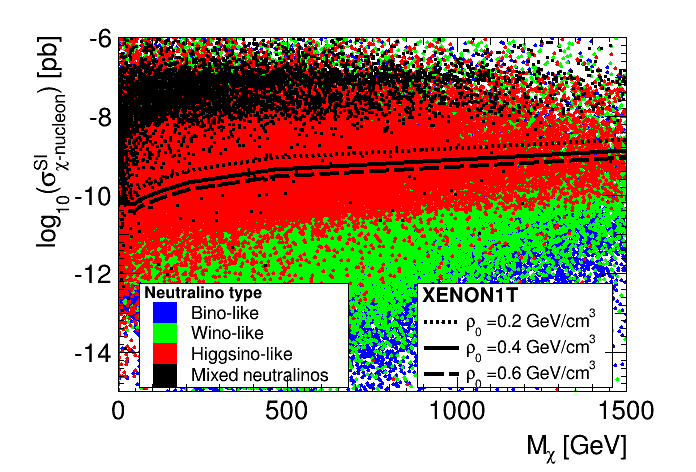} 
\caption{Generalised spin-independent neutralino scattering cross section as a function of the neutralino mass. The lines show the XENON1T 90\% C.L. upper limit for three different values of the local dark matter density $\rho_0$.\label{fig:DDSI_xenon}}
\end{center}
\end{figure}

In Fig.~\ref{fig:DDSI_xenon}, the generalised spin-independent \textsc{WIMP}-nucleon cross section -- which roughly corresponds to the $\chi$-xenon scattering cross section normalised to one nucleon, and which applies to xenon-based experiments -- is shown as a function of the neutralino mass, for the different neutralino 1 types. Higgsinos are in general more strongly interacting than the winos, leading to larger cross sections. In order to assess the consequences of the uncertainties on the obtained constraints, the recent limits of the XENON1T experiment are superimposed, for three values of the local dark matter density, namely $\rho_0 =$ 0.2, 0.4 and 0.6 GeV/cm$^3$. Between the conservative line corresponding to $\rho_0=0.2$ GeV/cm$^3$ and the most stringent limit obtained for $\rho_0=0.6$ GeV/cm$^3$, there is at most a factor 3 difference. While this is a large factor, in the context of pMSSM it does not change much the excluded region, which contains mainly Higgsino-like neutralinos 1.

\begin{figure}[t!]
\begin{center}
\includegraphics[width=\columnwidth]{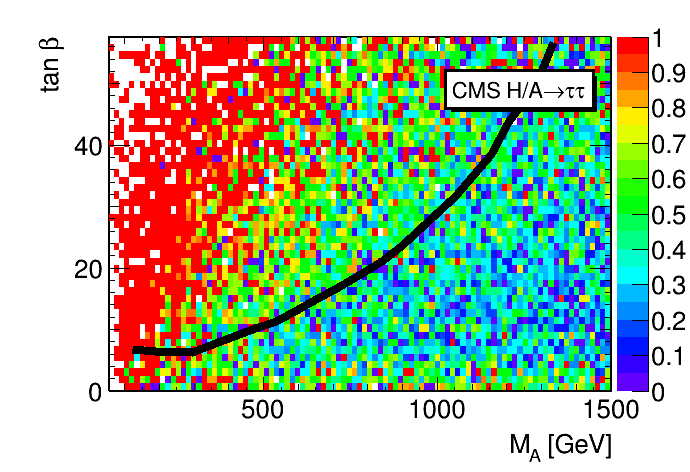} 
\caption{Fraction of points excluded by direct detection constraints in the $(M_A,\tan\beta)$ parameter plane. The CMS 13 TeV exclusion line from $H/A\to\tau\tau$ searches \cite{CMS:2016rjp} is also plotted for comparison.\label{DD_MAtanb}}
\end{center}
\end{figure}

In Fig.~\ref{DD_MAtanb} the exclusion by the XENON1T data with $\rho_0=0.4$ GeV/cm$^3$ is shown in the $(M_A,\tan\beta)$ parameter plane. For each bin, the fraction of excluded points is presented. This parameter plane is of interest since the neutral Higgs bosons can mediate the scattering, with couplings proportional to $\tan\beta$. About 100\% of the points are excluded in a triangle region starting from the origin of the plot and up to $\tan\beta=60$ and $M_A=600$ GeV. A large fraction of the points with larger $M_A$ can also be excluded. For comparison, the exclusion line from the CMS heavy Higgs searches for $H/A\to\tau\tau$ is also shown \cite{CMS:2016rjp}. While the CMS limit extends beyond the 100\% exclusion triangle and constitutes a well-defined and robust exclusion in this parameter plane, direct detection still adds complementary constraints for larger $M_A$ and smaller $\tan\beta$ values.

\begin{figure}[t!]
\begin{center}
\includegraphics[width=\columnwidth]{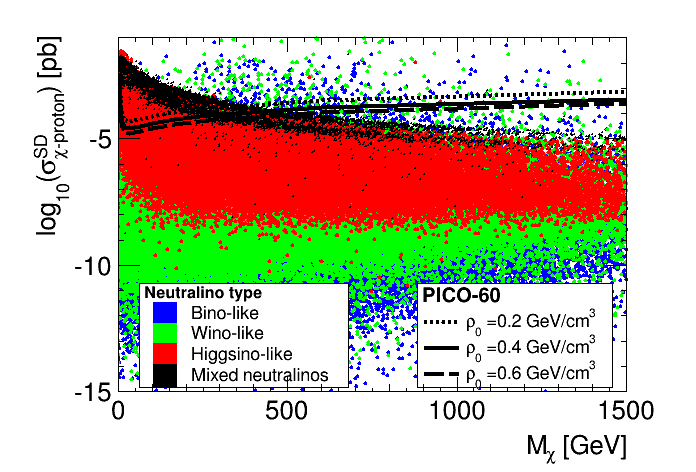}\\
\includegraphics[width=\columnwidth]{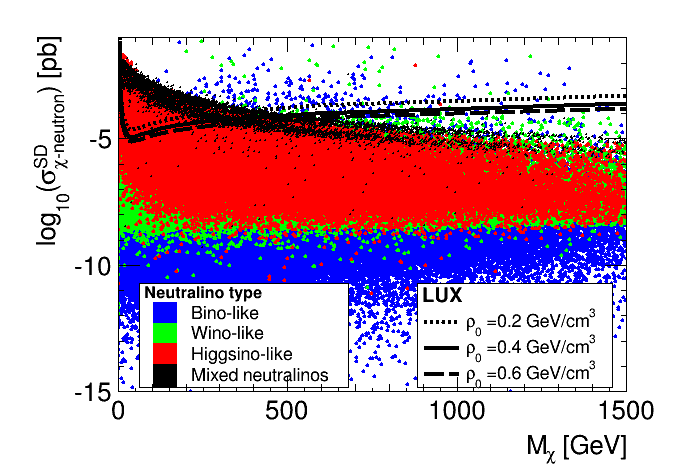} 
\caption{Spin-dependent neutralino scattering cross section with proton (upper panel) and with neutron (lower panel) as a function of the neutralino mass. The lines show the LUX and PICO-60 90\% C.L. upper limits for three different values of the local dark matter density $\rho_0$.\label{fig:DDSD}}
\end{center}
\end{figure}

LUX and PICO-60 also provide important constraints on the spin-dependent scattering cross section with protons and neutrons. This is shown in Fig.~\ref{fig:DDSD}, for $\rho_0$ between 0.2 and 0.6 GeV/cm$^3$. The distribution of the points is different for the proton and neutron scatterings, because the wino-neutralino 1 mixing term in the neutralino-quark-squark coupling is proportional to the isospin. In both cases however, only the most strongly interacting Higgsinos are excluded, and the value of $\rho_0$ does not affect much the results.

Practically, LUX and PICO-60 spin-dependent constraints are redundant, since both exclude the same points. The spin-independent XENON1T results give quite stringent constraints, which exclude most of the points probed by LUX and PICO-60. After imposing the XENON1T constraints, the spin-dependent results exclude about 0.5\% of the remaining points, with dominantly Higgsino-like $\chi$.

\subsubsection{Connections with relic density}

Direct detection constraints are not related to the relic density through annihilation cross sections, as for indirect detection. They are nevertheless complementary, since they provide constraints on different pMSSM parameters.

The same paradigm as for indirect detection can apply: if the relic density is smaller than the observed dark matter density, it may be because the neutralino is not the sole component of dark matter, thus the local dark matter density has to be rescaled accordingly to obtain the local neutralino density. As a consequence, the limits become less constraining, since the effective scattering cross sections are lowered by a factor proportional to the relic density. In comparison with indirect detection, the impact of the rescaling is less pronounced, because the rescaling is proportional to the dark matter density for direct detection, whereas it is proportional to the density squared for indirect detection.
 
\begin{table}[t!]
 \hspace*{-0.4cm}\begin{tabular}{|c|c|c|c|c|c|c|}
 \hline
  Neutralino&\multicolumn{2}{c|}{$\rho_{0}=0.2$}&\multicolumn{2}{c|}{$\rho_{0}=0.4$}&\multicolumn{2}{c|}{$\rho_{0}=0.6$}\\
    types&\multicolumn{2}{c|}{GeV/cm$^{3}$}&\multicolumn{2}{c|}{GeV/cm$^{3}$}&\multicolumn{2}{c|}{GeV/cm$^{3}$}\\
  \hline
  &No& With&No&With&No&With\\
 &Rescale&Rescale&Rescale&Rescale&Rescale&Rescale\\
 \hline
 Binos&$33.5$\%&$21.8$\%&$38.8$\%&$27.7$\%&$42.6$ \%&$31.9$\%\\
 Winos&$18.6$\%&$1.7$\%&$25.0$\%&$2.9$\%&$29.4$ \%&$3.7$\%\\
 Higgsinos&$50.2$\%&$12.1$\%&$63.2$\%&$18.1$\%&$71.1$ \%&$22.7$\%\\
 Mixed&$99.5$\%&$80.0$\%&$99.7$\%&$87.0$\%&$99.8$ \%&$89.9$\%\\
 \hline
  \textbf{All}&$33.5$\%&$8.8$\%&$42.2$\%&$12.1$\%&$47.7$ \%&$14.3$\%\\
 \hline
 \end{tabular}
 \caption{Fraction of points, valid after imposing the relic density upper limit, that are excluded by direct detection limits, for the different neutralino types. The exclusions are set for different values of the local DM density, which is rescaled or not by the relic density.\label{tab:DD_exclusion}}
\end{table}

In Table~\ref{tab:DD_exclusion}, the fractions of excluded points are given for the different neutralino 1 types, with rescaling and without rescaling, for $\rho_0=0.2,0.4,0.6$ GeV/cm$^3$, after the upper limit of the relic density is applied. First, in absence of rescaling, even in the most conservative case corresponding to $\rho_0=0.2$ GeV/cm$^3$, direct detection imposes strong limits, and one third of the points are excluded. The Higgsinos are the most affected, followed by the binos and winos. The mixed states are almost completely excluded by direct detection, but their number is too small to draw statistically significant conclusions. When increasing the density to $\rho_0=0.6$ GeV/cm$^3$, the sensitivity is enhanced, with about half of the points excluded, and 70\% of the Higgsinos. With the relic density rescaling, the exclusion power decreases strongly, as only 15\% of the points remain excluded in the most favourable case. The exclusion hierarchy is also modified in presence of rescaling, with the binos being the most excluded neutralinos 1.

\subsubsection{Combined dark matter constraints}

Dark matter observables can lead to very strong constraints. The relic density, if compared to both upper and lower bounds of the measured dark matter density, leads to a very strong exclusion and a selection of the points with small mass splittings or resonant annihilations. However, the relic density constraint suffers from uncertain hypotheses about the Early Universe, which prevents us from considering the lower bound. Applying only the upper bound, most of the bino-like $\chi$ which are not accompanied by other SUSY particles close in mass are excluded.
Indirect detection brings complementary constraints, and probes wino-like and Higgsino-like neutralinos. We showed that the different assumptions on the galactic halo profiles and propagation models for the cosmic rays can however modify the limits on the cross sections by a few orders of magnitude and strongly lower the constraining power of indirect detection limits.
Direct detection on the other hand sets strong constraints on the pMSSM, with an exclusion of 25-40\% of our scan points, depending on the local dark matter density. The main uncertainty is due to the local dark matter density, but even if the conservative choice lowers the exclusion power, it does not affect much the excluded parameter region.

It is however important to recall that in the case the neutralino 1 is not responsible for the whole dark matter density and a rescaling of the dark matter density is necessary for direct and indirect detections, dark matter observables severely lose their constraining power. In the following, we focus on the case where neutralinos constitute the whole dark matter and do not consider the scaling possibility any further. 

\begin{figure*}[t!]
\begin{tabular}{ccc}
  \includegraphics[width=.33\linewidth]{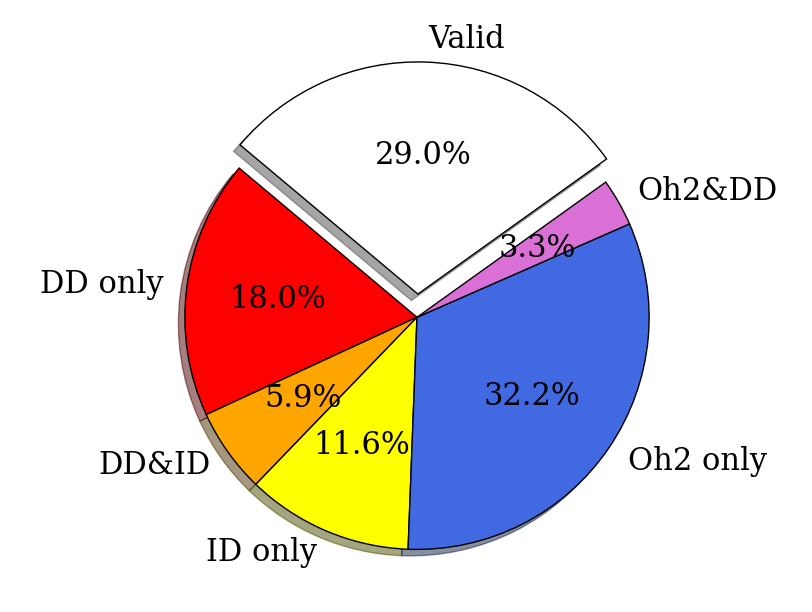}&
  \includegraphics[width=.33\linewidth]{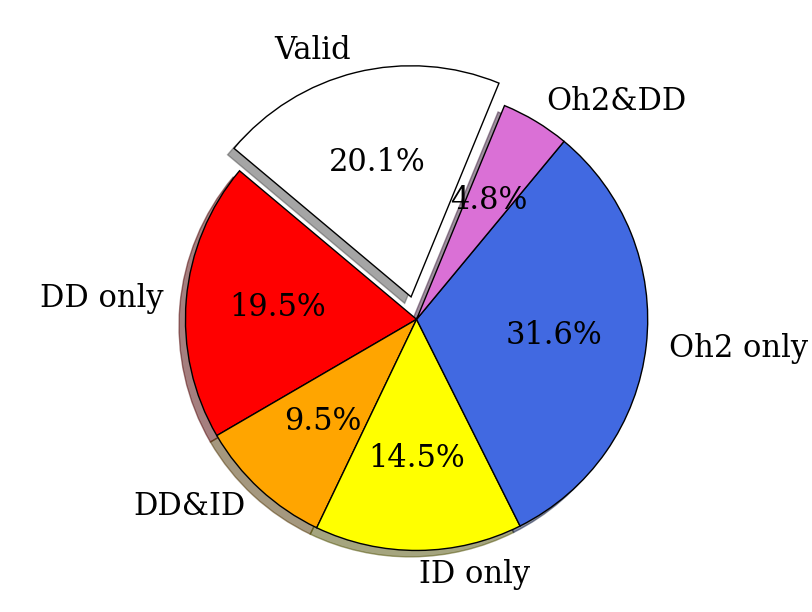}&
  \includegraphics[width=.33\linewidth]{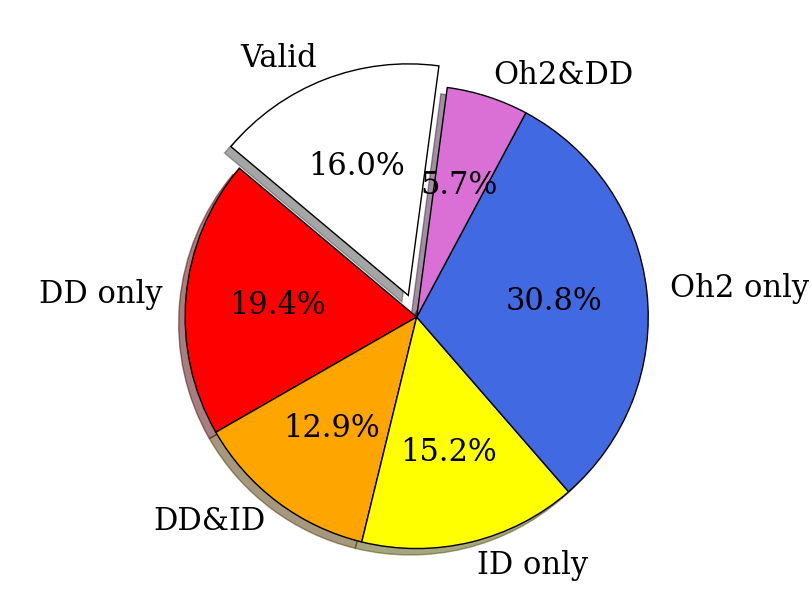}\\
   CONSERVATIVE&STANDARD&STRINGENT\\
   \end{tabular}
 \caption{Fraction of pMSSM points excluded by upper bound of the dark matter density, direct detection and indirect detection constraints.\label{pie:allDM}}
\end{figure*}

We will now quantitatively study the interplay between the different dark matter constraints. We define three cases:
\begin{itemize}
 \item CONSERVATIVE: $\rho_0=0.2$ GeV/cm$^3$ for direct detection, Burkert dark matter profile and cosmic ray \MED~propagation model using AMS-02 data for indirect detection.
 \item STANDARD: $\rho_0=0.4$ GeV/cm$^3$ for direct detection, NFW dark matter profile using the combined analysis of the 19 confirmed dwarf spheroidal galaxies observed by Fermi-LAT for indirect detection.
 \item STRINGENT: $\rho_0=0.6$ GeV/cm$^3$ for direct detection, Einasto dark matter profile and cosmic ray \MAX~propagation model using AMS-02 data for indirect detection.
\end{itemize}

In Fig.~\ref{pie:allDM}, the fraction of pMSSM points initially satisfying the light Higgs mass constraint, which are excluded by the upper bound of the dark matter density, direct detection and indirect detection constraints, is shown for the three cases of astrophysical assumptions. The \& symbol corresponds to the exclusive ``and''. Points excluded simultaneously by the relic density and indirect detection constraints represent less than 1\% of the total number of points, and are not shown.

The relic density constraint excludes about 36\% of the points. As already seen, direct detection constraints are relatively insensitive to the choice of the local density of dark matter, and direct detection excludes 25\% of the points in the conservative case and 35\% in the stringent case. Indirect detection is more sensitive to the choice of profile and propagation model and excludes less than 20\% of the points in the conservative case and 30\% in the stringent one. In all cases, the simultaneous application of the dark matter constraints is very important, and allows us to strongly reduce the number of valid points, even in the most conservative case. 

\begin{figure*}
\centering
\begin{tabular}{ccc}
&\underline{\bf BINO}&\\
  \includegraphics[width=.33\linewidth]{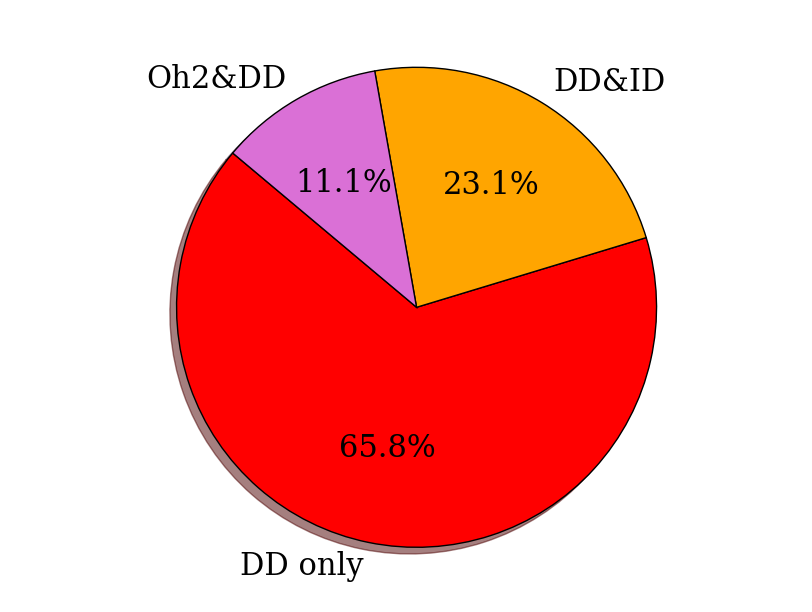}&
  \includegraphics[width=.33\linewidth]{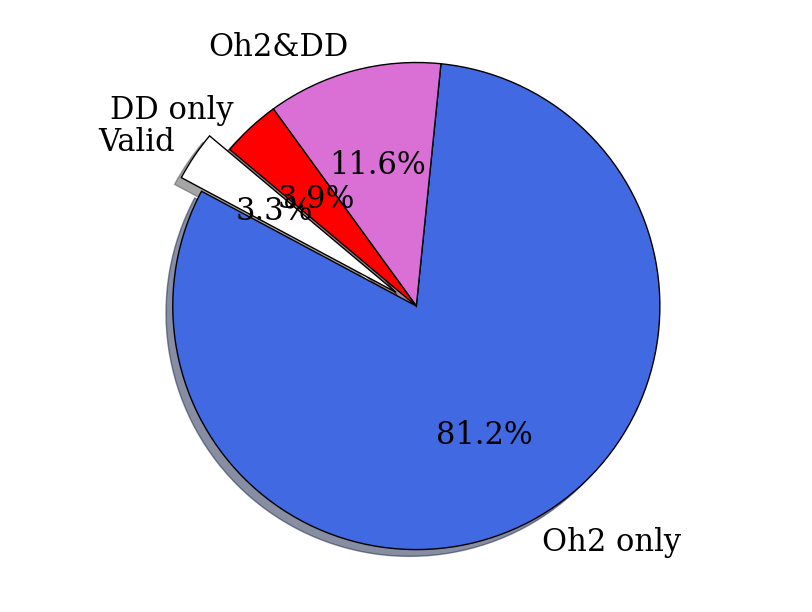}&
  \includegraphics[width=.33\linewidth]{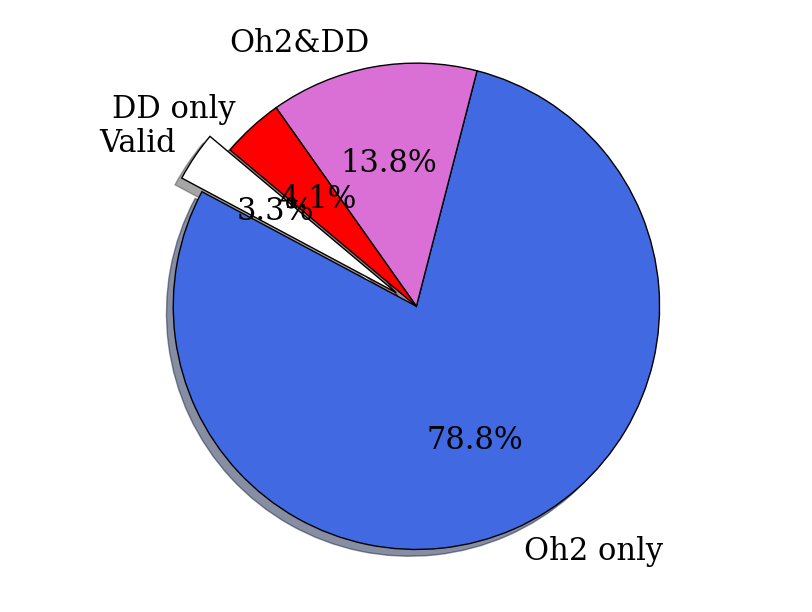}\\
  CONSERVATIVE&STANDARD&STRINGENT\\
\hline
   \end{tabular}
\begin{tabular}{ccc}
&\underline{\bf WINO}&\\
  \includegraphics[width=.33\linewidth]{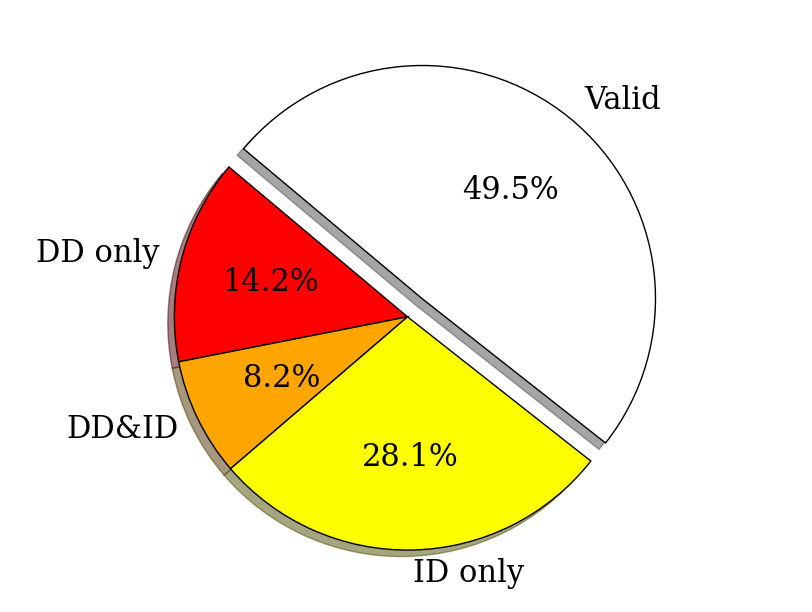}&
  \includegraphics[width=.33\linewidth]{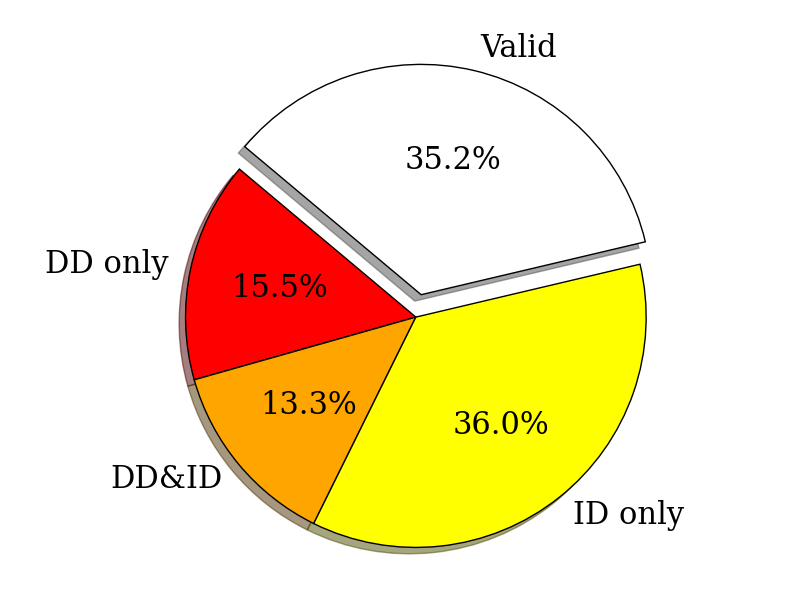}&
  \includegraphics[width=.33\linewidth]{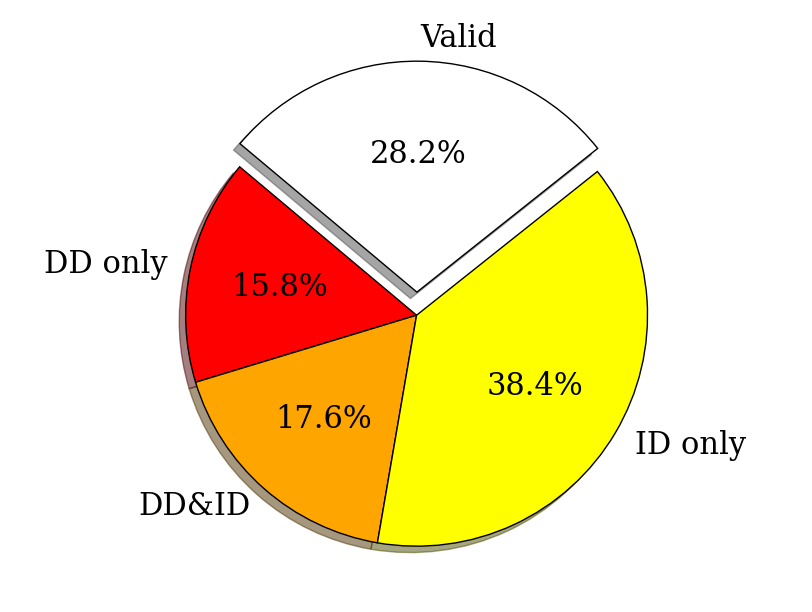}\\
   CONSERVATIVE&STANDARD&STRINGENT\\
\hline
   \end{tabular}
\begin{tabular}{ccc}
&\underline{\bf HIGGSINO}&\\
  \includegraphics[width=.33\linewidth]{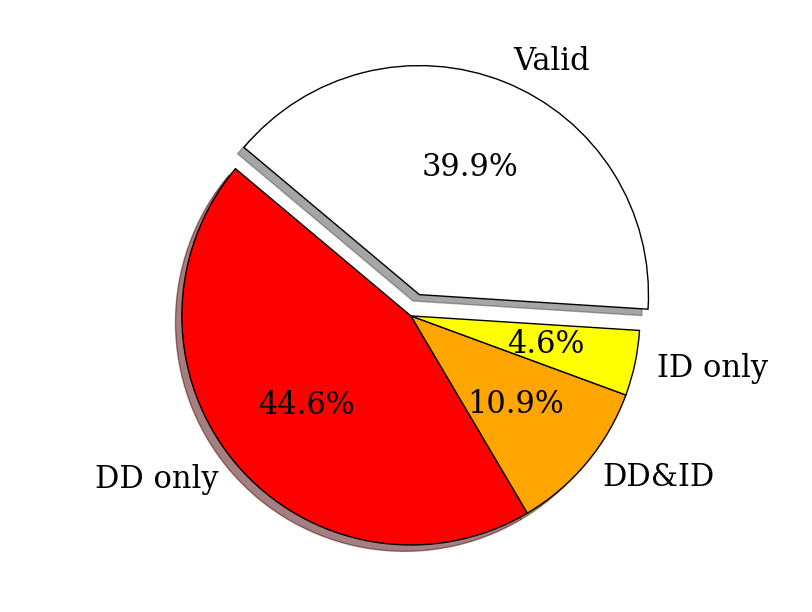}&
  \includegraphics[width=.33\linewidth]{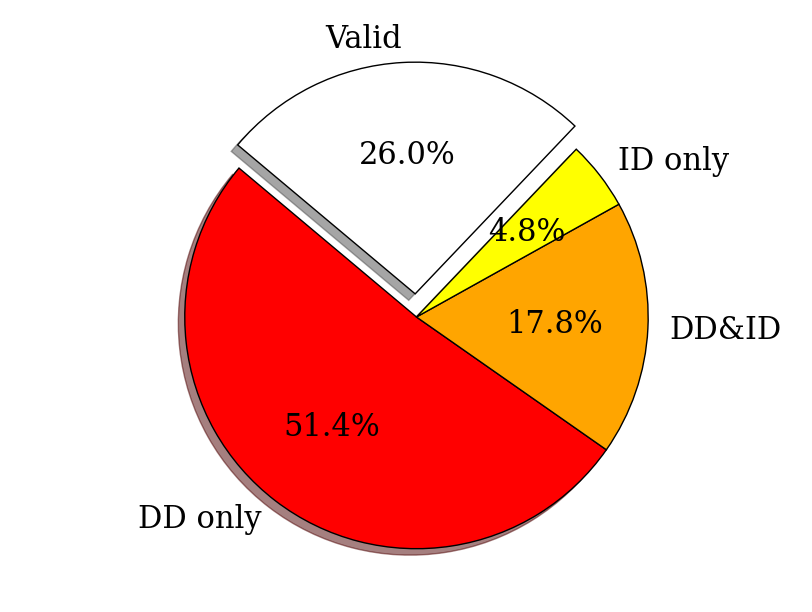}&
  \includegraphics[width=.33\linewidth]{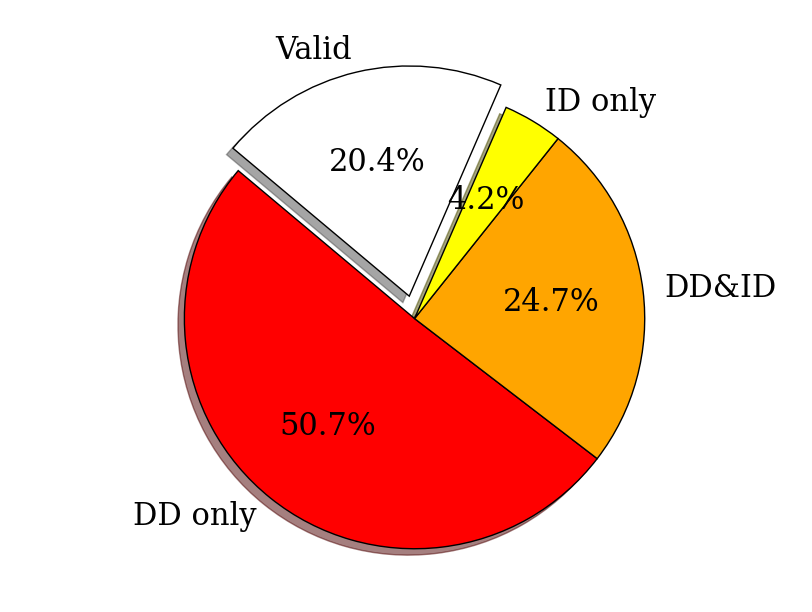}\\
   CONSERVATIVE&STANDARD&STRINGENT\\
\hline
   \end{tabular}
\begin{tabular}{ccc}
&\underline{\bf MIXED}&\\
  \includegraphics[width=.33\linewidth]{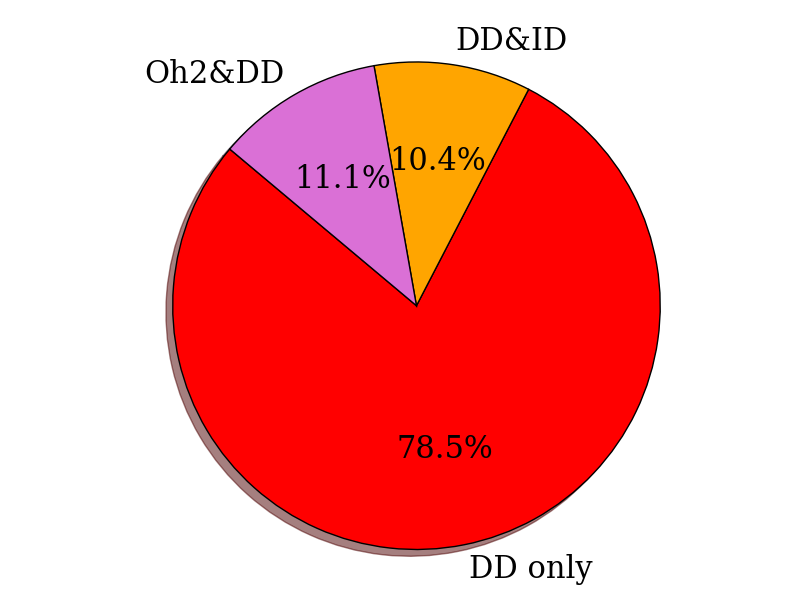}&
  \includegraphics[width=.33\linewidth]{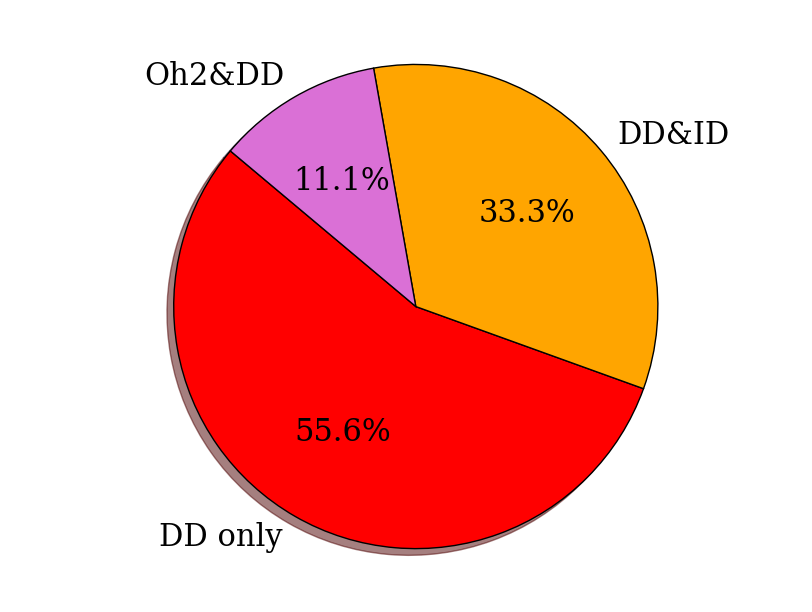}&
  \includegraphics[width=.33\linewidth]{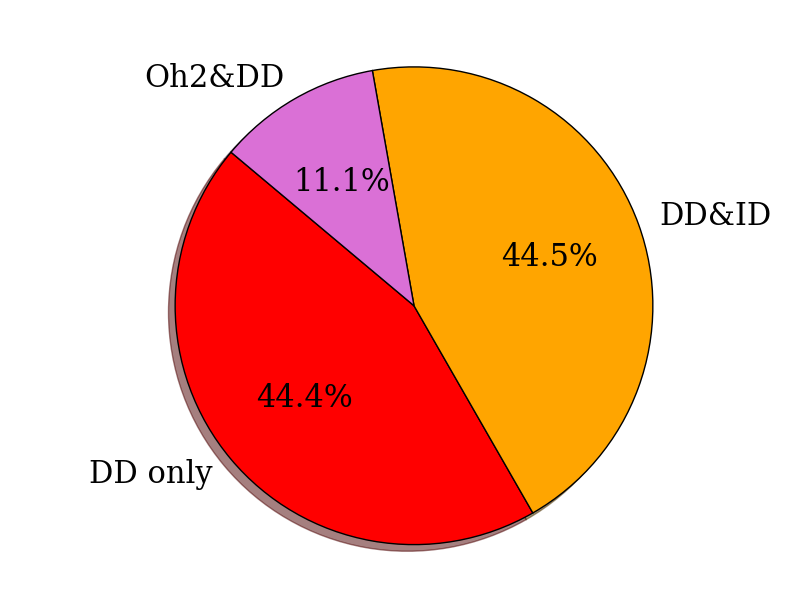}\\
   CONSERVATIVE&STANDARD&STRINGENT\\
   \end{tabular}
\caption{Fraction of pMSSM points excluded by the upper bound of the dark matter density, direct detection and indirect detection constraints for the different neutralino 1 types.\label{pie:typesDM}}
\end{figure*}

In Fig.~\ref{pie:typesDM}, the same analysis is performed for the different neutralino types separately. 
First, the bino-like neutralinos 1 have in general weaker couplings, leading to large relic densities and small annihilation and scattering cross sections. Thus, the bino-like points are strongly excluded by the relic density, slightly probed by direct detection, and negligibly by indirect detection. Therefore, the choice of the conservative or stringent constraints has a negligible effect, since the exclusion is dominated by the relic density.
Second, wino-like neutralinos 1 are dominantly excluded by indirect detection, followed by direct detection. After these constraints, relic density only affects a negligible fraction of points, which is why the exclusion by relic density does not appear in the figure. For the winos, the choice of the conservative or stringent cases strongly affects the results, leaving 50\% of the points valid in the conservative case, and 28\% in the stringent case. Again, the standard case leads to results similar to the stringent case.
Third, the Higgsino-like neutralinos 1 are mainly excluded by direct detection, which mildly depends on the astrophysical hypotheses. Indirect detection also excludes a number of points, even if a large fraction of them is already excluded by direct detection. As for the winos, relic density only excludes a negligible fraction of points after the direct and indirect detection constraints. At the end, 40\% of the Higgsinos remain valid in the conservative case, and 20\% in the stringent case.
Finally, the mixed-state neutralinos 1 are completely excluded independently from the astrophysical hypotheses, and predominantly by direct detection.\\
\\
To summarise this section, dark matter constraints set strong constraints on the pMSSM parameter space. However, while direct detection leads to relatively robust constraints, indirect detection is more sensitive to the choice of galaxy halo profiles and cosmic ray propagation models.

\subsection{Collider and Dark Matter constraints}
\label{sec:mssm-lhc}

In this section, the complementarity of collider and dark matter constraints will be studied.

Whereas dark matter limits are subject to astrophysical and cosmological uncertainties or hypotheses, collider constraints are obtained in environments under control, which therefore lead to relatively hypothesis-free limits.

As explained in Section~\ref{sec:collider}, using the LHC results requires the computation of numerous cross sections, generation of events and detector simulation, which are computationally heavy and CPU-time consuming. In order to gain CPU time, we perform the event generation and detector simulation only for model points which respect the light Higgs mass constraint, flavour  physics, and LEP and Tevatron constraints, as well as the upper bound of the relic density. The points satisfying these constraints will be referred to as ``Accepted points'' in the following.

\begin{figure}[t!]
  \centering
  \includegraphics[width=0.6\linewidth]{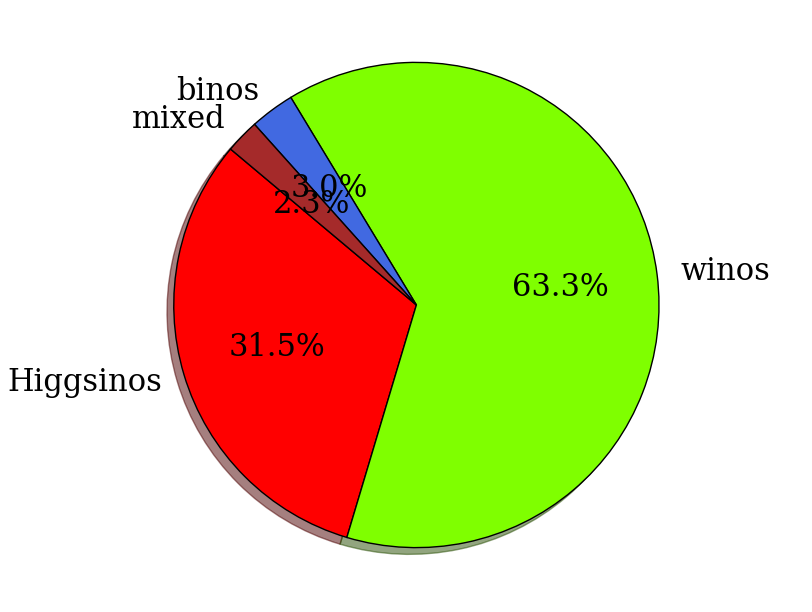}
  \caption{Fractions of neutralino 1 types in our scan after imposing the light Higgs mass limit, LEP and flavour constraints, and relic density upper bound.\label{pie:prelhc}}
\end{figure}

In Fig~\ref{pie:prelhc}, we present the type of neutralinos 1 for the accepted points. A comparison with Fig.~\ref{pie:initial} showing the type of the points satisfying only the light Higgs mass limit, reveals that most of the binos have been excluded, but that the fraction of winos in comparison with the Higgsinos is unchanged. This is mainly due to the upper bound of the relic density, as explained in Section~\ref{sec:mssm-relic}. The LEP and flavour constraints do not probe directly the neutralino 1, but can affect scenarios with light wino-like and Higgsino-like $\chi$ through the constraints on the charginos and heavier neutralinos. Nevertheless, the exclusion power of these constraints is limited in comparison to the relic density one.

In the Higgs sector, the light Higgs mass constraint favours the decoupling limit where the heavy Higgs bosons are heavy, and heavy stop masses with maximal mixing \cite{Djouadi:2005gj,Arbey:2011ab,Bechtle:2012jw,CahillRowley:2012rv}. Measurements of the light Higgs production and decay channels also point towards large heavy Higgs masses. In particular, the diphoton channel favours heavy charginos, stops and charged Higgs bosons \cite{Djouadi:2005gj,Carena:2011aa,Carena:2012gp,Benbrik:2012rm,Arbey:2012dq}. In addition, light Higgs decays into supersymmetric particles are rather limited \cite{Djouadi:2005gj,Arbey:2012bp,Djouadi:2012rh,Chakraborti:2014gea,Chakraborti:2017dpu}. These important limits provide strong constraints in the ($\mu,M_2$) parameter plane. Indeed, both parameters are important for the neutralino and chargino mixings, and $\mu$ is also important for the third generation squark mixings. The limits obtained from the measurements of the light Higgs couplings are complemented by the electroweakino direct searches at LEP and the LHC.
\begin{figure}[t!]
\begin{center}
\includegraphics[width=\columnwidth]{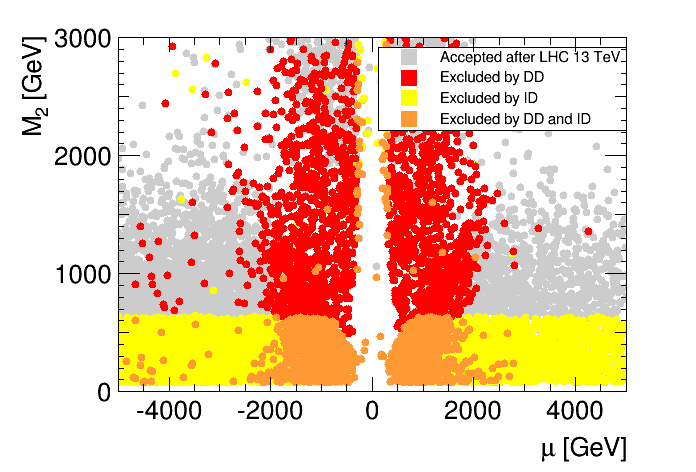} 
\caption{pMSSM points in the ($\mu,M_2$) parameter plane. The accepted parameter points which are in agreement with the LHC 8 and 13 TeV data from Higgs and direct searches are shown in gray. The red points are in addition excluded by direct detection, the yellow points by indirect detection and the orange points by direct and indirect detections simultaneously.\label{fig:MuM2}}
\end{center}
\end{figure}
This is illustrated in Fig.~\ref{fig:MuM2}, where the small $\mu$ values are excluded. The complementarity with dark matter constraints is rather clear. Direct detection excludes points spread over the plane. Indirect constraint severely excludes points with $M_2 \lesssim 600$ GeV and $|\mu| \lesssim 150$ GeV. One should however note that due to the multi-dimensional parameter space, there could be points below the coloured regions that still survive the dark matter and collider constraints.

\begin{figure}[t!]
\begin{center}
\includegraphics[width=\columnwidth]{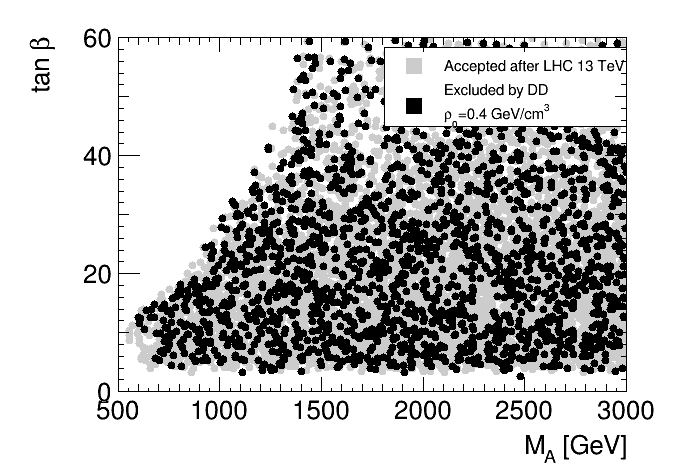} 
\caption{pMSSM points in the ($M_A,\tan\beta$) parameter plane. The accepted model points which are in agreement with the LHC 8 and 13 TeV data from Higgs and direct searches are shown in gray. The black points are excluded by direct detection.\label{lhc_MAtanb}}
\end{center}
\end{figure}

The heavy Higgs searches, and in particular $H/A\to\tau\tau$ searches, impose strong constraints in the ($M_A,\tan\beta$) parameter plane which is also relevant for direct detection as seen in Fig.~\ref{DD_MAtanb}. In Fig.~\ref{lhc_MAtanb}, we superimpose over the points in agreement with the LHC constraints those which are excluded by direct detection. Similarly to direct detection, $H/A\to\tau\tau$ searches probe the large $\tan\beta$ and small $M_A$ region (corresponding to the empty region in the upper right part in the figure). We can see from the figure that the exclusion by direct detection is not well defined and spread. Comparing with Fig.~\ref{DD_MAtanb} reveals that the strongest and well defined exclusion by direct detection in this plane occurs below the $H/A\to\tau\tau$ limit. Both constraints are nevertheless complementary and allow us to exclude points beyond the large $\tan\beta$ and small $M_A$ region.

\begin{figure}[t!]
\begin{center}
\includegraphics[width=\columnwidth]{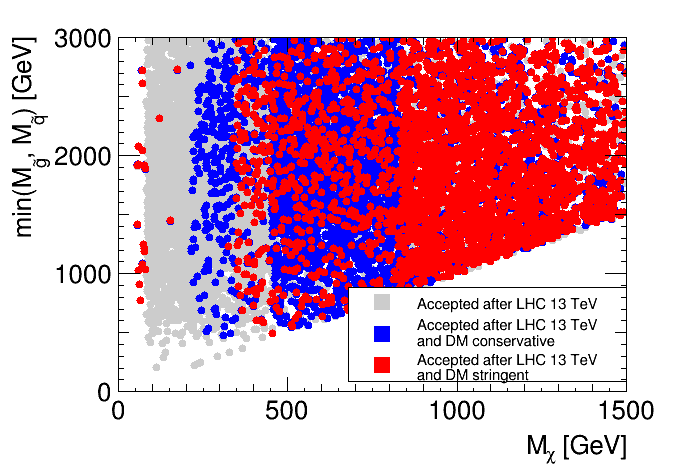} 
\caption{pMSSM points in the ($M_\chi,M_{\tilde{g},\tilde{q}}$) parameter plane. $M_{\tilde{g},\tilde{q}}$ is the lightest mass among the gluino and first and second generation squark masses. The accepted parameter points which are in agreement with the LHC 8 and 13 TeV data from Higgs and direct searches are shown in gray. The points which in addition agree with dark matter constraints with conservative astrophysical hypotheses are in blue, and with stringent hypotheses in red.\label{lhc_MN1MGL}}
\end{center}
\end{figure}

\begin{figure}[t!]
\begin{center}
\includegraphics[width=\columnwidth]{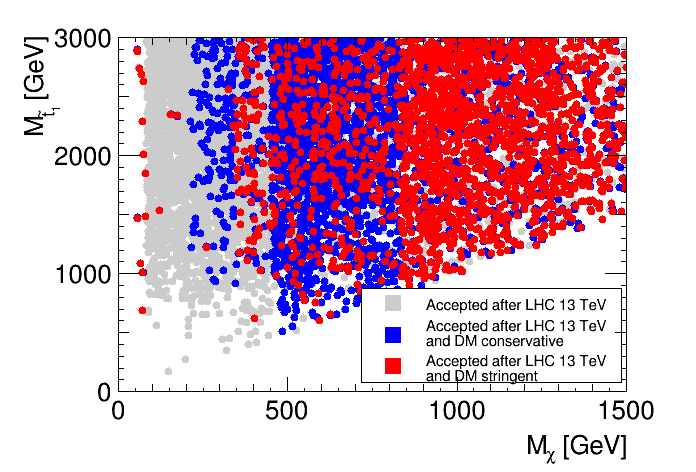} 
\caption{pMSSM points in the ($M_\chi,M_{\tilde{t}_1}$) parameter plane. The accepted parameter points which are in agreement with the LHC 8 and 13 TeV data from Higgs and direct searches are shown in gray. The points which in addition agree with dark matter constraints with conservative astrophysical hypotheses are in blue, and with stringent hypotheses in red.\label{lhc_MN1MSTOP}}
\end{center}
\end{figure}

\begin{figure*}
\begin{center}
\begin{tabular}{ccc}
  \includegraphics[width=.33\linewidth]{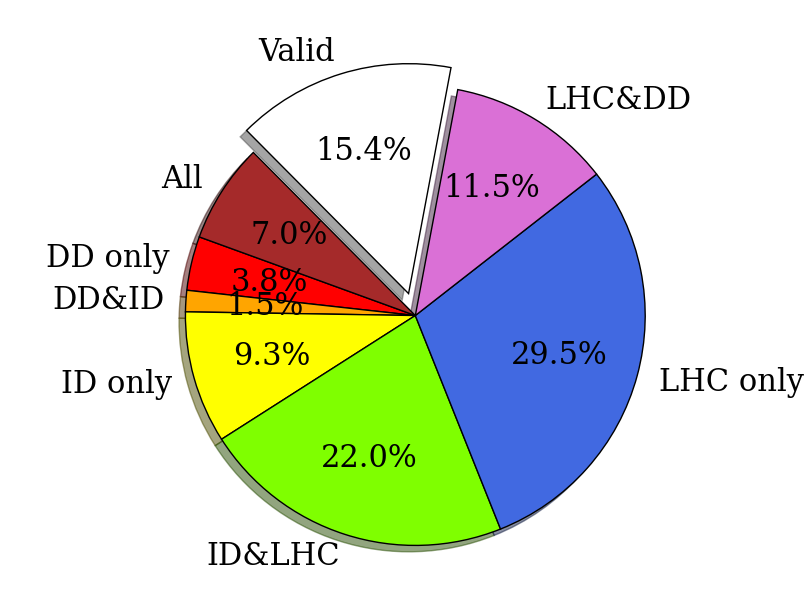}&
  \includegraphics[width=.33\linewidth]{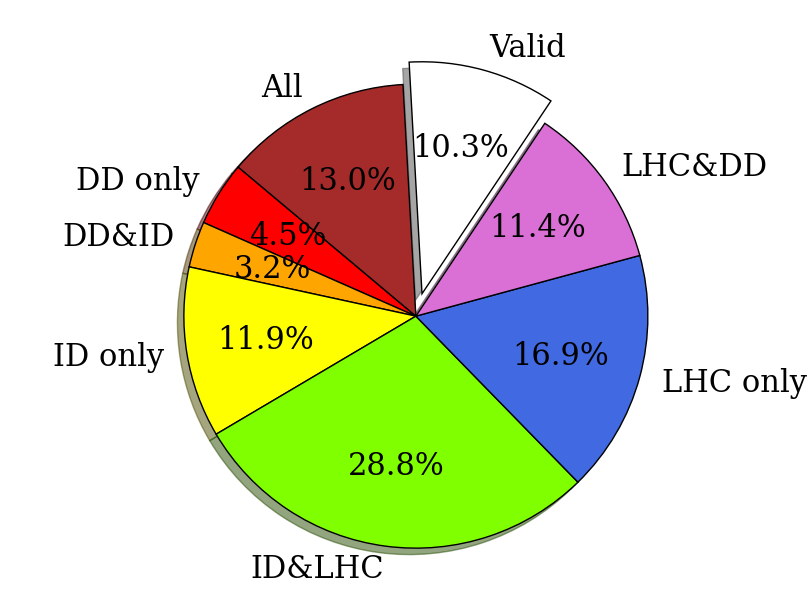}&
  \includegraphics[width=.33\linewidth]{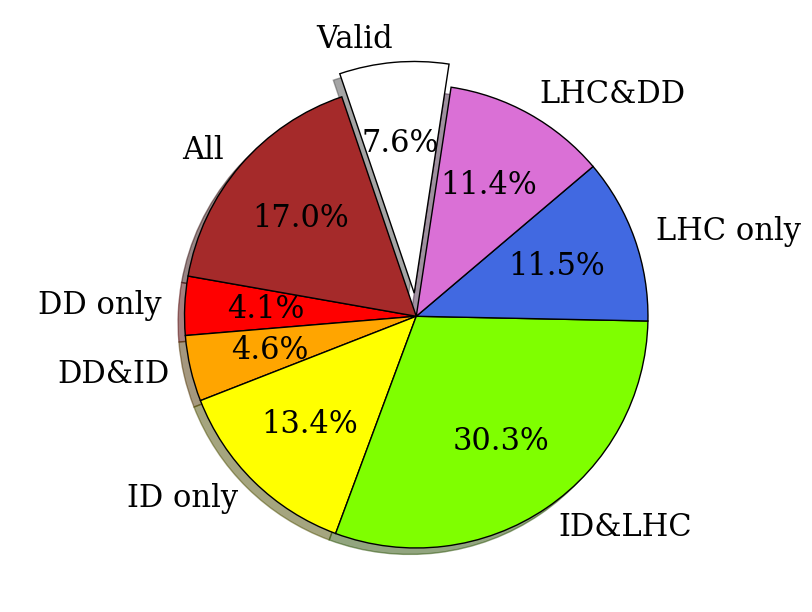}\\
   CONSERVATIVE&STANDARD&STRINGENT\\
   \end{tabular}
 \caption{Fraction of the pMSSM points satisfying the light Higgs mass, relic density, LEP and flavour constraints excluded by direct and indirect detections and LHC constraints.\label{pie:allDMLHC}}
 \end{center}
\end{figure*}

As a hadron collider, LHC is more sensitive to strongly interacting particles. In particular, gluinos and squarks of the first and second generations are amongst the most actively searched particles, and LHC can probe masses as large as a few TeV in the most favourable scenarios. In Fig.~\ref{lhc_MN1MGL}, the accepted pMSSM points are plotted in the minimum mass amongst the gluino and first and second generation squark masses vs. neutralino 1 mass plane. We note that gluinos or squarks as light as a few hundred GeV can still escape LHC searches in a general scenario as the pMSSM. These points correspond mainly to compressed scenarios \cite{LeCompte:2011cn,LeCompte:2011fh,Bhattacherjee:2012mz,Dreiner:2012gx}, where one or more supersymmetric particles have masses close-by, leading to decays with particles or jets in the final state which can leave the detectors undetected because of their small transverse energies. Dark matter searches can be very important in these cases and exclude points which are not probed at the LHC, as can be seen from the figure. Direct detection probes points spread over the plane. Indirect detection can probe neutralino 1 masses up to 450 GeV in the conservative case, 800 GeV in the stringent case, independently of the squark and gluino masses. We also see that after the LHC constraints, light squarks or gluinos of a few hundred GeV in compressed or complicated scenarios are still allowed, but after the dark matter constraints, they are less numerous and the surviving points correspond to very small squark/gluino-neutralino 1 mass splittings, and in the stringent case the squark and gluino masses are pushed beyond 450 GeV. So the complementarity is obvious, as dark matter experiments can probe parameter regions which are not accessible at the LHC, and vice versa.

Similar result for the lightest stop is presented in Fig~\ref{lhc_MN1MSTOP}. As for the gluino and squark case, light stops are still allowed by collider constraints in compressed scenarios, which can still be probed by dark matter detection experiments. Light stop scenarios which escape LHC detection are still allowed, but the stop 1 mass is pushed beyond 500 GeV in the conservative case and 600 GeV in the stringent case, after imposing the direct and indirect detection limits.

Finally, the interplay of the LHC and dark matter constraints is presented in a quantitative way in Fig.~\ref{pie:allDMLHC}. It can be seen that the LHC has the major role in probing the pMSSM parameter space, but dark matter detection constraints further probe the parameter space. The combination of all constraints leads to an exclusion of between 85\% and 92\% of our sample.

\begin{figure*}
\centering
\begin{tabular}{ccc}
&\underline{\bf BINO}&\\
  \includegraphics[width=.33\linewidth]{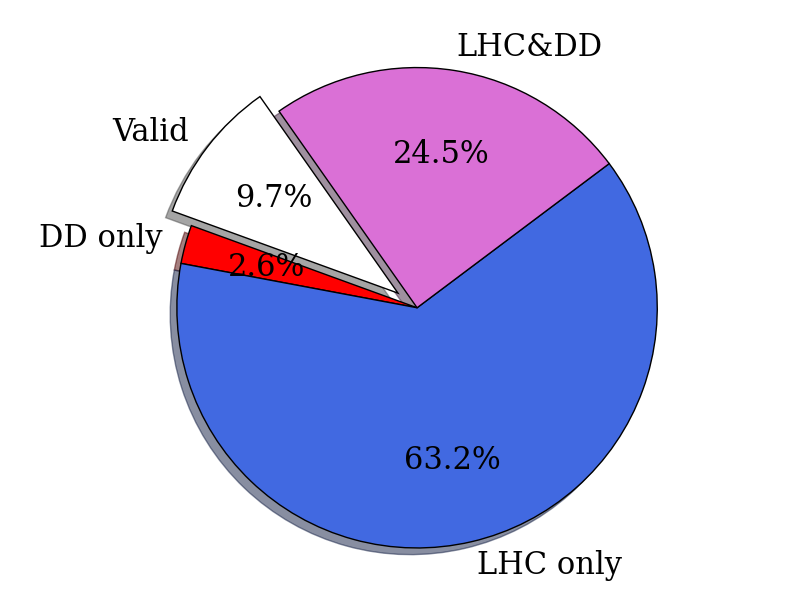}&
  \includegraphics[width=.33\linewidth]{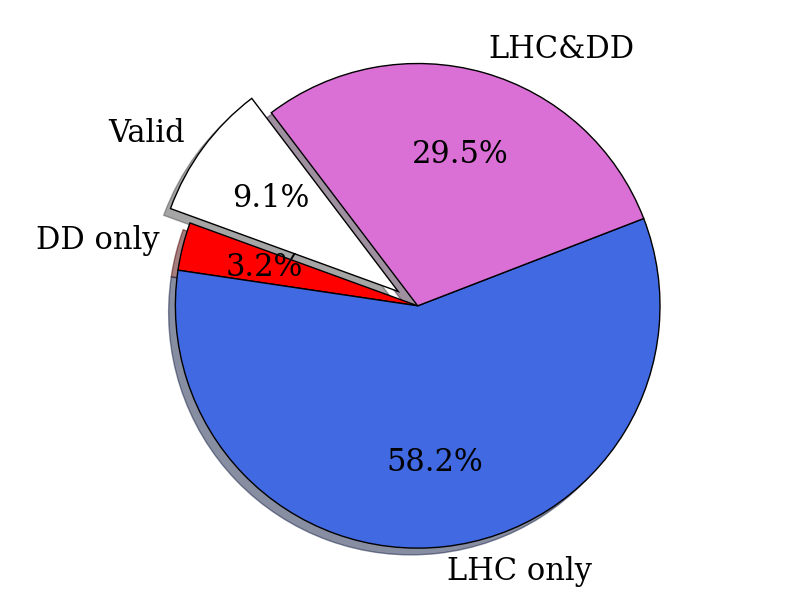}&
  \includegraphics[width=.33\linewidth]{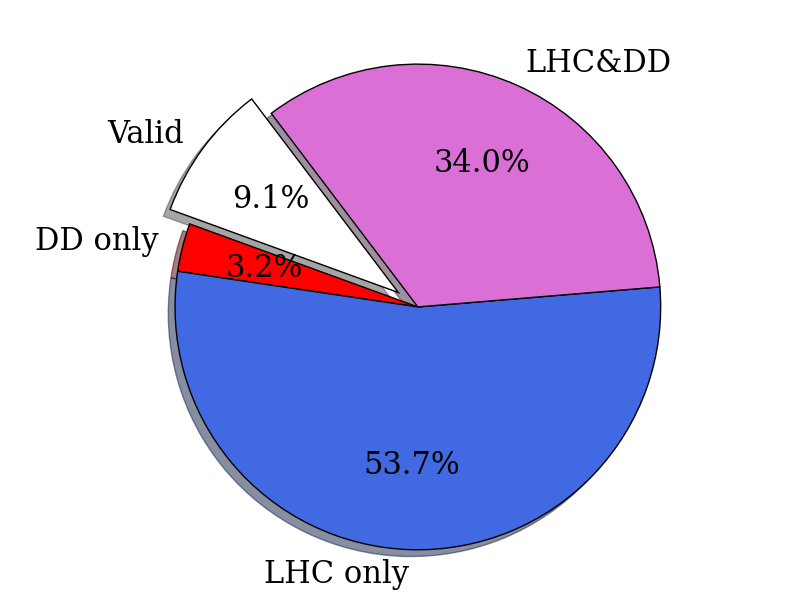}\\
   CONSERVATIVE&STANDARD&STRINGENT\\
\hline
   \end{tabular}
\begin{tabular}{ccc}
&\underline{\bf WINO}&\\
  \includegraphics[width=.33\linewidth]{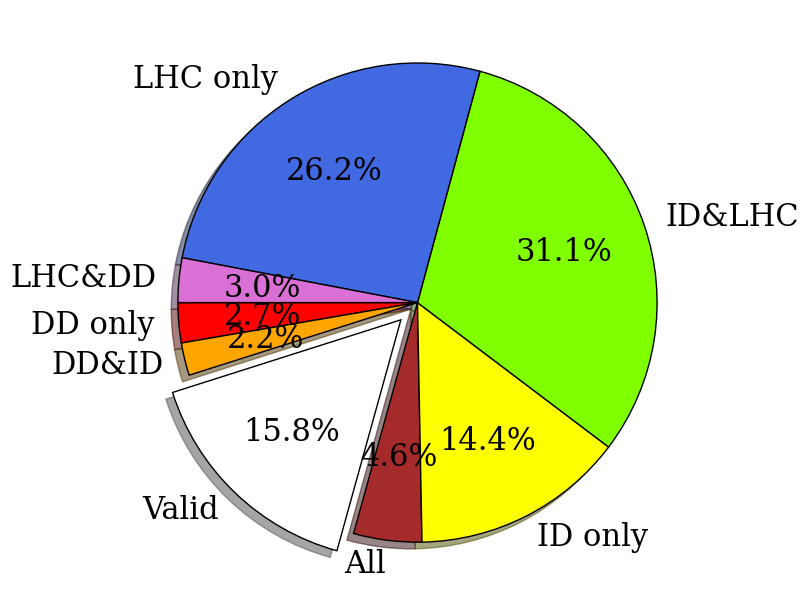}&
  \includegraphics[width=.33\linewidth]{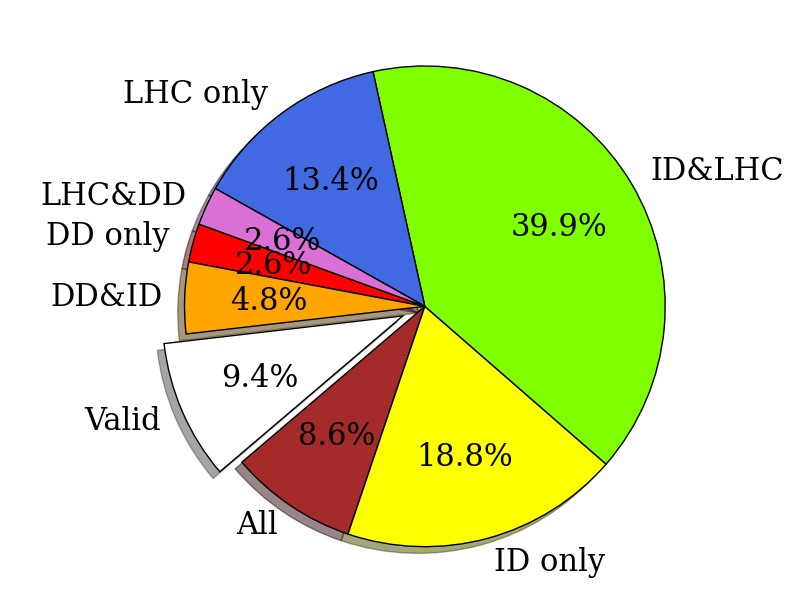}&
  \includegraphics[width=.33\linewidth]{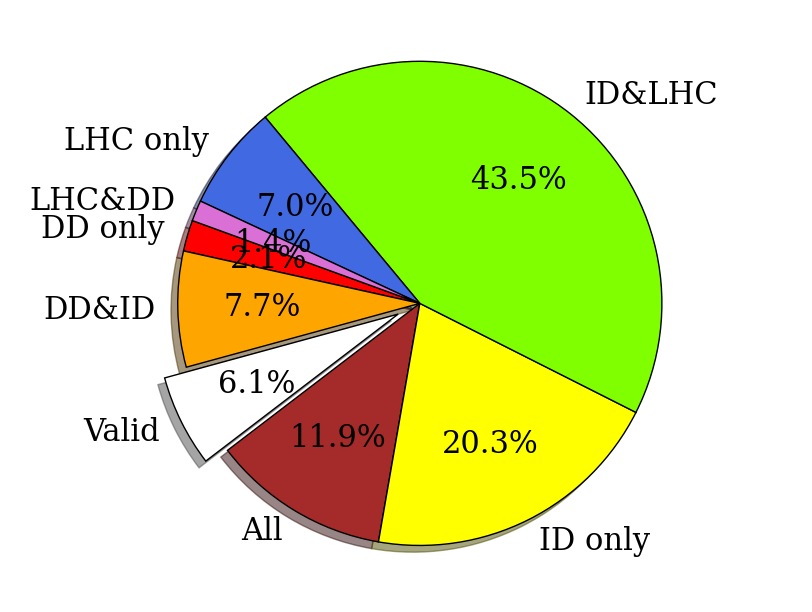}\\
   CONSERVATIVE&STANDARD&STRINGENT\\
\hline
   \end{tabular}
\begin{tabular}{ccc}
&\underline{\bf HIGGSINO}&\\
  \includegraphics[width=.33\linewidth]{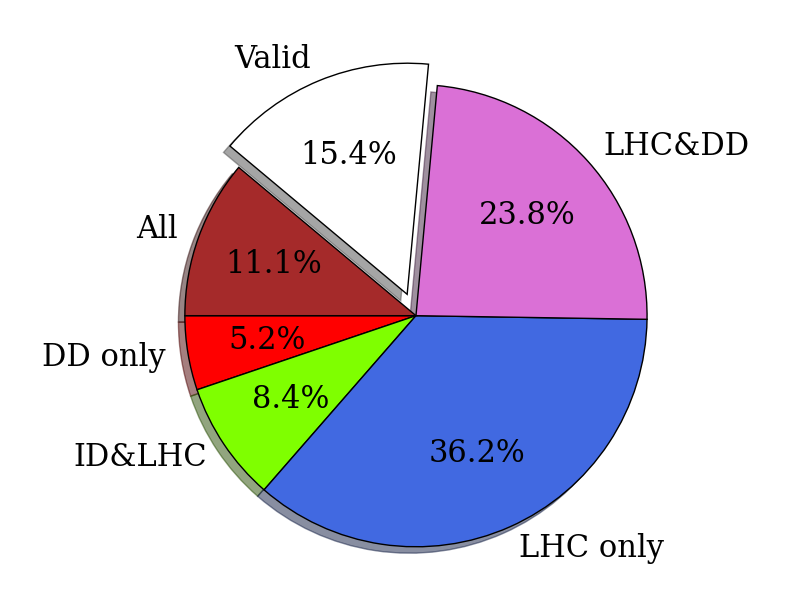}&
  \includegraphics[width=.33\linewidth]{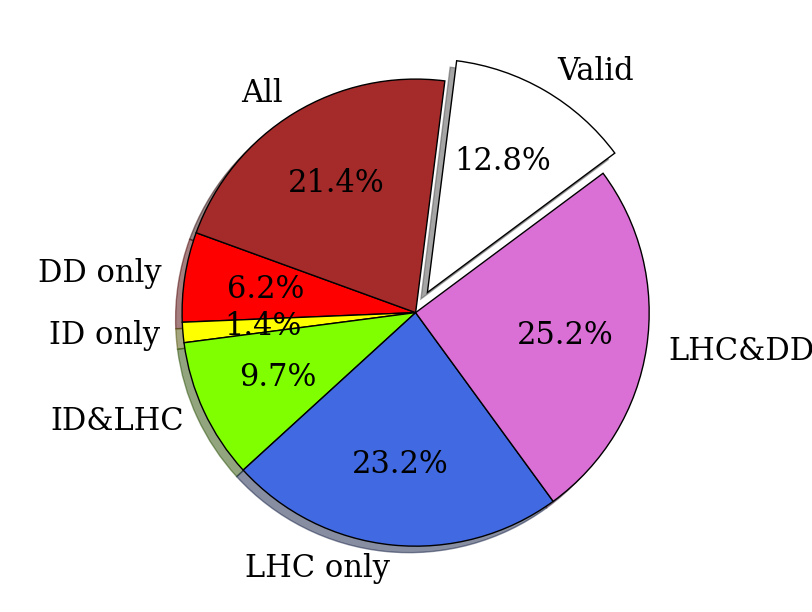}&
  \includegraphics[width=.33\linewidth]{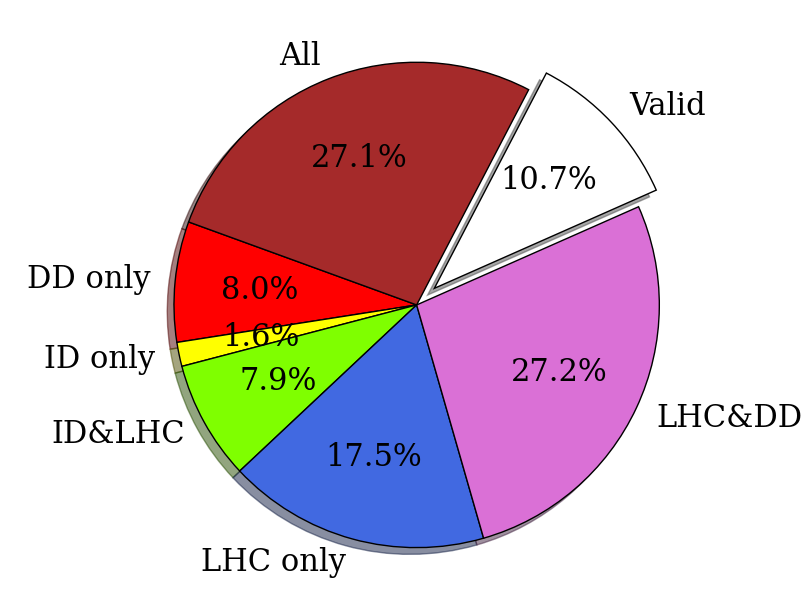}\\
   CONSERVATIVE&STANDARD&STRINGENT\\
\hline
   \end{tabular}
\begin{tabular}{ccc}
&\underline{\bf MIXED}&\\
  \includegraphics[width=.33\linewidth]{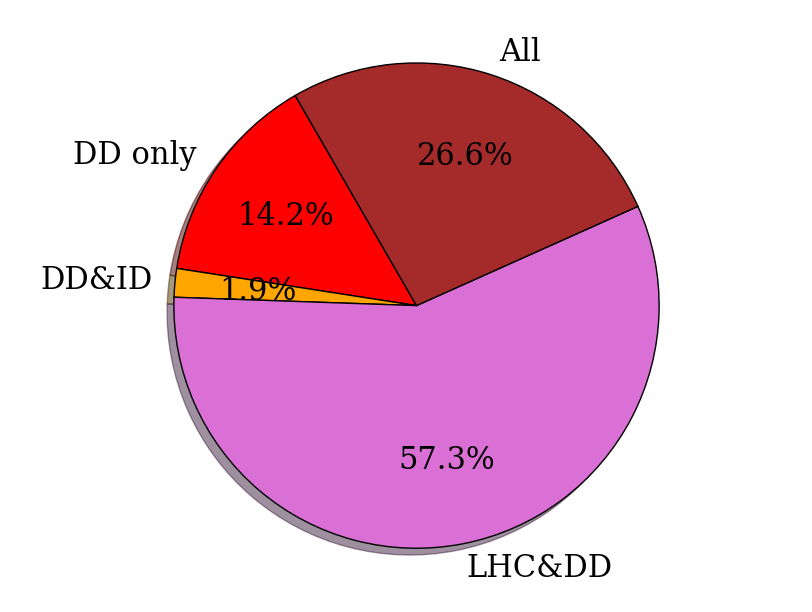}&
  \includegraphics[width=.33\linewidth]{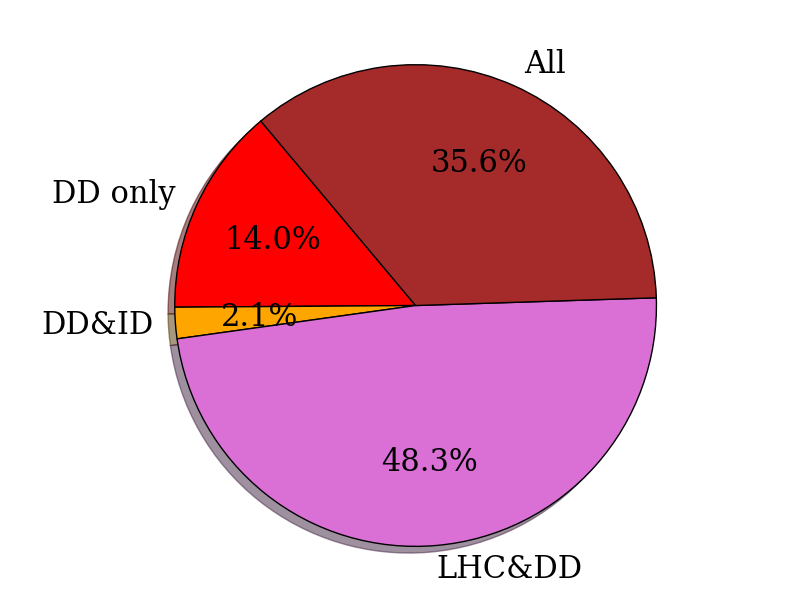}&
  \includegraphics[width=.33\linewidth]{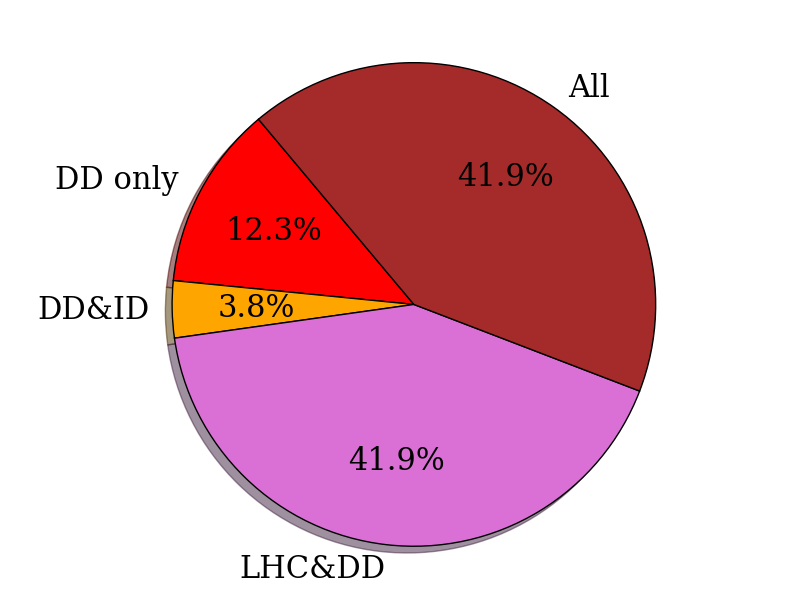}\\
   CONSERVATIVE&STANDARD&STRINGENT\\
   \end{tabular}
 \caption{Fraction of pMSSM points satisfying the light Higgs mass, relic density, LEP and flavour constraints, and excluded by direct and indirect detections and LHC constraints, for the different neutralino 1 types.\label{pie:typesDMLHC}}
\end{figure*}
 
Fig.~\ref{pie:typesDMLHC} presents a more detailed view of the exclusion for the different neutralino 1 types. In particular, it reveals that LHC excludes more than 65\% of the points independent of the neutralino 1 type. The role of dark matter constraints on the contrary is more type-dependent. As we showed earlier, binos, Higgsinos and mixed states are more strongly probed by direct detection, while indirect detection rather excludes winos. And whereas direct detection is mildly sensitive to the choice of the astrophysical parameters, indirect detection is more sensitive to it.

\begin{figure*}
\begin{tabular}{ccc}
  \includegraphics[width=.33\linewidth]{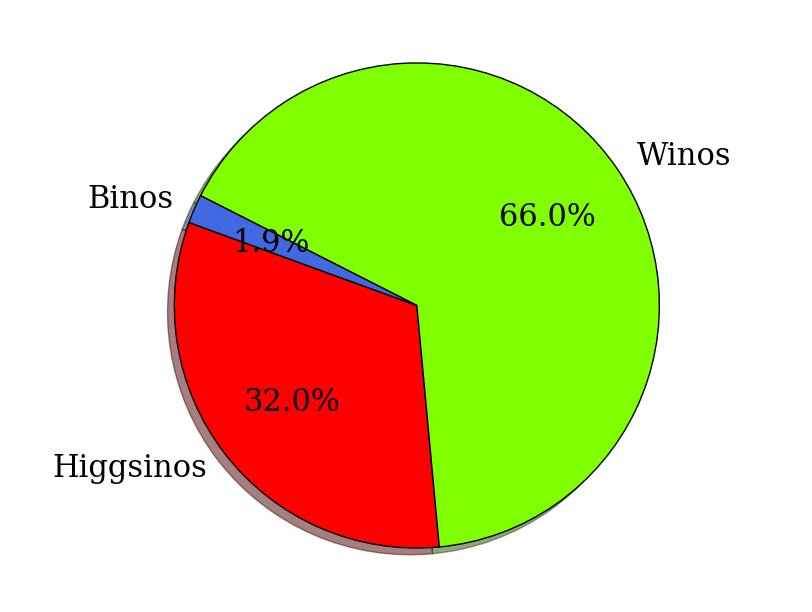}&
  \includegraphics[width=.33\linewidth]{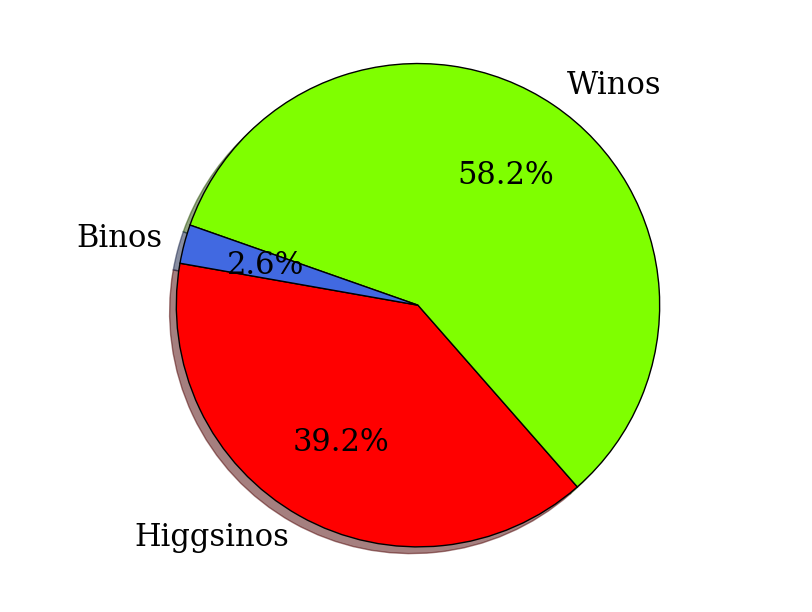}&
  \includegraphics[width=.33\linewidth]{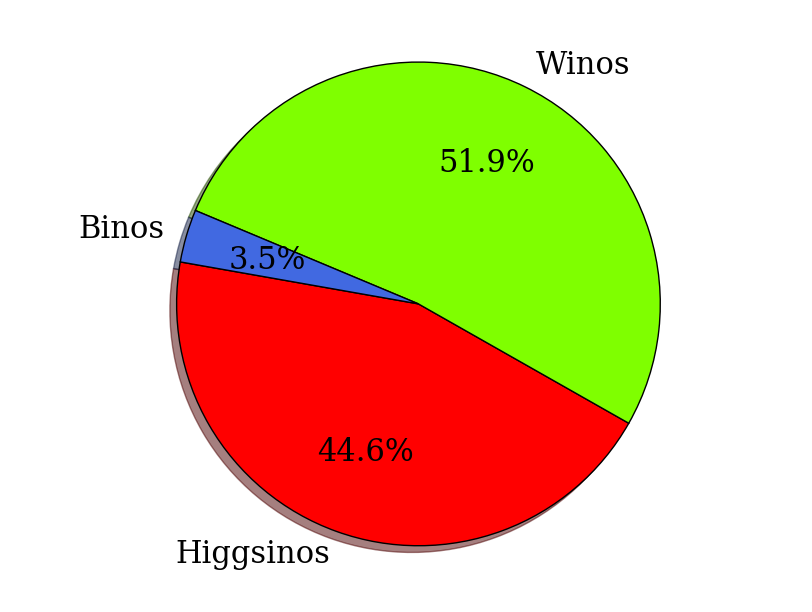}\\
   CONSERVATIVE&STANDARD&STRINGENT\\
   \end{tabular}
 \caption{Fractions of neutralino 1 types in our scan after imposing all the constraints (including only the upper bound for the relic density).\label{pie:allconstraints}\vspace*{0.3cm}}
\end{figure*}
\begin{figure*}
\begin{tabular}{ccc}
  \includegraphics[width=.33\linewidth]{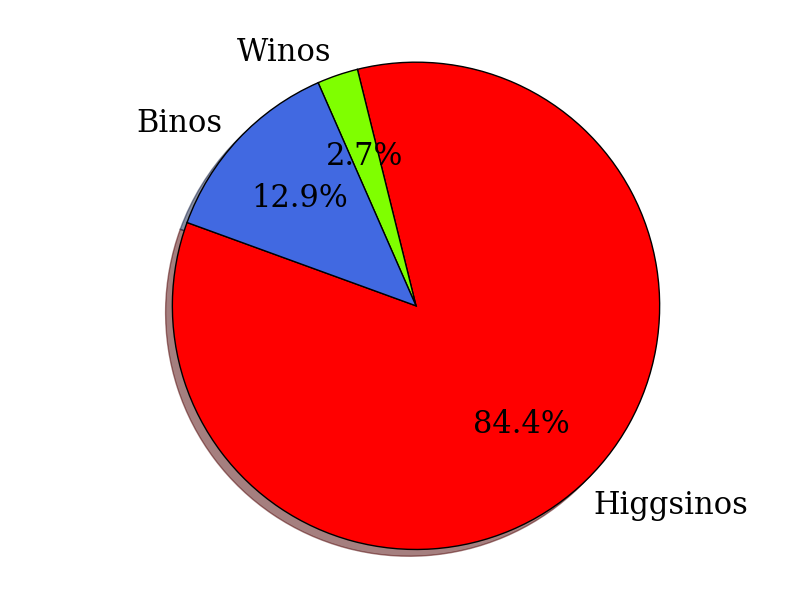}&
  \includegraphics[width=.33\linewidth]{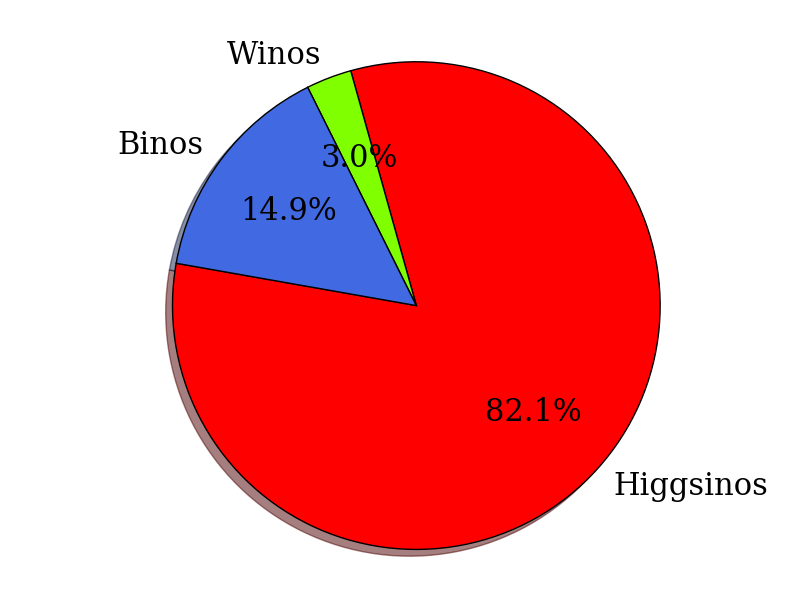}&
  \includegraphics[width=.33\linewidth]{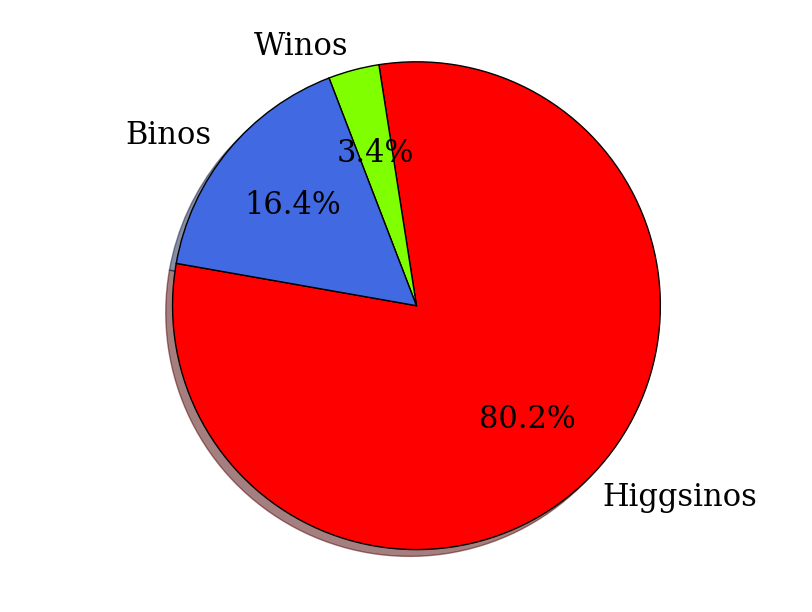}\\
   CONSERVATIVE&STANDARD&STRINGENT\\
   \end{tabular}
 \caption{Fractions of neutralino 1 types in our scan after imposing all the constraints, including also the lower relic density limit.\label{pie:allconstraintsinf}\vspace*{0.3cm}}
\end{figure*}

Finally, in Fig.~\ref{pie:allconstraints}, the fraction of neutralino 1 types after imposing all the constraints is shown. This figure is to be compared with Fig.~\ref{pie:allDMLHC}, where only LEP, flavour and relic density constraints were applied. We can see that the final fractions are still similar after applying all constraints, with a larger proportion of winos, followed by a large proportion of Higgsino, and a small amount of binos. This shows that the relic density constraint is the most type-selecting constraint. However, we note that the proportion of winos is much larger in the conservative dark matter case than in the stringent case.

An important caveat here is in order. The fraction of points has no real statistical meaning, but rather shows the tendency of the constraints to select certain types. To illustrate this, we show in Fig.~\ref{pie:allconstraintsinf} the fraction of the types after applying all the constraints, including the Planck lower bound. In this case, the Higgsinos are now the dominant surviving species, followed by the binos, and the winos survive only in small proportion. It is interesting to note that in this case, the choice of conservative or stringent astrophysical hypotheses does not affect much the results.\\
\\

\begin{figure}[t!]
\begin{center}
\includegraphics[width=\columnwidth]{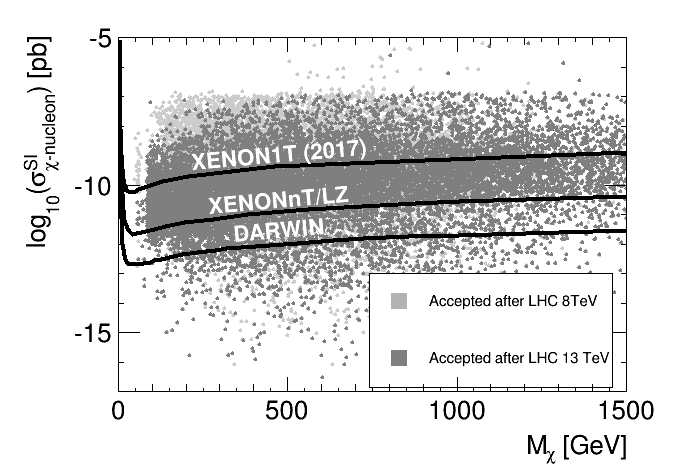} 
\caption{pMSSM points in the spin-independent scattering cross section vs. neutralino 1 mass parameter plane. The current XENON1T upper limit is superimposed together with the prospective limits of XENONnT/LZ and DARWIN.\label{DD_future}}
\end{center}
\end{figure}

Before concluding, it is worth mentioning that great improvements in the sensitivity of the direct and indirect detection experiments are expected in the coming years. Concerning direct detection, in the next few years XENONnT \cite{Aprile:2015uzo} and LZ \cite{Akerib:2015cja} will push the XENON1T limit by two orders of magnitude, and within ten years DARWIN \cite{Aalbers:2016jon} will allow us to gain one extra order of magnitude. This is illustrated in Fig.~\ref{DD_future}. For comparison, the gray points correspond to a sample of our points which are in agreement with the current LHC 8 TeV and 13 TeV limits. Practically, XENONnT/LZ will exclude most of the Higgsino points, and DARWIN will be able to probe a large part of the wino region. In addition, we have shown that the constraining power of direct detection is only mildly affected by the choice of the astrophysical assumptions, thus these limits will provide relatively robust constraints on the pMSSM parameter space. The DARWIN limit will however be close to the neutrino background, which constitutes a large obstacle to further improvements. Nevertheless, the remaining points after DARWIN will have mainly wino-like neutralinos 1, which will be probed by indirect detection.

\begin{figure}[t!]
\begin{center}
\includegraphics[width=\columnwidth]{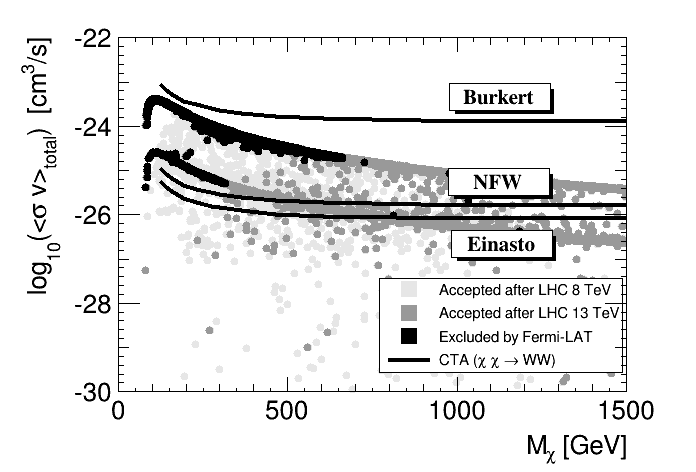} 
\caption{Total annihilation cross section as a function of the neutralino 1 mass. The CTA prospective upper limits are superimposed for the Einasto, NFW and Burkert profiles.\label{ID_future}\vspace*{0.3cm}}
\end{center}
\end{figure}

For indirect detection, the Cherenkov Telescope Array (CTA) \cite{Carr:2015hta}, dedicated to gamma rays, will use a Cherenkov imagery technique similar to HESS, VERITAS or MAGIC, and will be able to probe an energy range between a few tenths of GeV to above 100 TeV. Before 2030, CTA will also further push the indirect detection limits by observing gamma rays at the center of the Milky Way, as shown in Fig.~\ref{ID_future}. It is important to remark however, that contrary to the Fermi-LAT limits, which are obtained from the observations of spheroidal dwarves and which are therefore less affected by the dark matter profile, since CTA will focus on the galaxy center, it is subject to strong uncertainties from the dark matter profile. 
Since the question of the existence of cuspy profiles is unresolved \cite{Silk:2016srn}, dark matter density distributions such as NFW or Einasto which incorporate cuspy profiles, will lead to fundamentally different exclusion limits than a Burkert profile with a core.
This is illustrated in the figure, a Burkert profile will lead to limits which are two orders of magnitude less constraining than the NFW or Einasto profile. Therefore, CTA will be even more subject to astrophysical uncertainties, even if we can hope for an improvement of our knowledge of the galactic center within the next decade.

%% file: conclusions.tex
In this paper, we studied the impact of dark matter direct and indirect detections, in conjunction with relic density and collider constraints, on the phenomenological MSSM with neutralino dark matter and addressed in some detail the consequences of the related uncertainties.

First, the calculation of the relic density is based on the assumption of radiation dominating the Early Universe properties. Any deviation from this hypothesis can modify the relic density by orders of magnitude. In addition, dark matter may be made of different components, the neutralino being only one of them. These considerations justify the usage of solely the upper relic density bound. In the pMSSM, applying both relic density bounds selects compressed scenarios with co-annihilations or with annihilation through a $Z$-boson or Higgs boson resonance, and favours bino-like neutralinos 1. Disregarding the lower bound strongly changes this picture, and rather selects Higgsino- and wino-like neutralinos 1.

We then reviewed the calculation of indirect detection constraints, which relies on the choice of a dark matter halo profile for gamma rays, as well as a propagation model for cosmic rays. We showed that between the more conservative case, corresponding to the Burkert halo and the \MED~propagation model, and the most stringent case, Einasto profile and \MAX~propagation model, the limits from AMS-02 differ by one order of magnitude. In the context of pMSSM, indirect detection excludes more strongly the wino-like neutralinos 1, followed by the Higgsino-like ones. 

Turning to direct detection, we showed that the constraints are affected by the local relic density in the vicinity of the Earth, whose evaluations can vary within a factor 3, and to a lesser extent by the velocity of Earth in the dark matter halo. As a consequence, the exclusion limits obtained by direct detection experiments can vary by a factor 3. Even if this uncertainty is rather large, in the context of pMSSM, it does not strongly affect the excluding power of direct detection. Independent of the choice of the local density, direct detection is particularly efficient in probing scenarios with Higgsino-like neutralino 1.

An interesting aspect of the connection of direct and indirect detections with relic density comes from the possibility that dark matter could be made of several components. In such a case, the neutralino relic density is smaller than the measured dark matter density. Therefore, the local and galactic neutralino dark matter densities are smaller than the measured ones, and have to be rescaled by the ratio of the relic density over the dark matter density measured by Planck. Such a rescaling strongly alleviates the indirect detection constraints since they are proportional to the density squared, and decreases to a lesser extent the direct detection limits, which are proportional to the density.

Apart from this specific case, direct detection, indirect detection and (the lower bound of) relic density are very efficient in the pMSSM, excluding a large part of our sample. Even with the most conservative choice of astrophysical uncertainties, 70\% of our sample is excluded by the dark matter constraints, and 85\% in the most stringent case. As expected, constraints from indirect detection are the ones which are the most affected by the astrophysical assumptions, but the complementarity between the dark matter constraints is still of major importance when studying BSM scenarios.

We then studied the interplay of dark matter constraints with collider constraints from LEP, Tevatron, $B$-factories and LHC searches. The LHC is by design more apt to probe the strong interaction sector than the weak interaction one. On the contrary, dark matter searches probe the weak interaction sector. As a consequence, there is an interesting complementarity between the two kinds of searches. The Higgs boson however is also related to the weak sector. In addition, flavour physics and heavy Higgs searches allow us to explore the chargino and heavy Higgs sectors. 

There exists two parameter planes where the complementarity between collider and dark matter constraints is obvious: the ($M_A,\tan\beta$) plane, which is probed by BR($B_s\to\mu^+\mu^-$) in flavour physics, $H/A\to\tau\tau$ searches at the LHC and dark matter direct detection. We showed that, while heavy Higgs searches provide very robust constraints in this parameter plane, dark matter direct detection can provide less strict constraints which spread beyond those of the LHC. The second plane is ($M_2,\mu$), which is very important for the light Higgs and chargino sectors. LEP and LHC searches for electroweakinos and Higgs results allowed us to exclude small $|\mu|$ and $M_2$, but direct and indirect detections further probe this parameter region up to $M_2 \lesssim 600$ GeV and $|\mu| \lesssim 150$ GeV.

Furthermore, the LHC excludes strongly gluino and squarks of intermediate masses, but is less sensitive to scenarios with compressed spectra, which lead to invisible final states. As a consequence, light squarks or gluinos of a few hundreds of GeV can still escape detection. We showed that dark matter constraints exclude these light gluinos and squarks, including light stops. This highlights the importance of the complementarity between dark matter and collider constraints.

Finally, the latest collider constraints alone exclude 70\% of our sample of points, and dark matter constraints alone between 55\% and 80\% depending on the astrophysical assumptions. Altogether, the exclusion reaches between 85\% and 93\% of our scan points, showing again the complementary of the collider and dark matter experiments, regardless of the astrophysical hypotheses.

In the future, there will be interesting prospects for dark matter direct and indirect detections. In particular, in the coming years XENONnT and LZ will improve the current limits by two orders of magnitude, and DARWIN within ten years by one extra order of magnitude. Similarly, the gamma ray telescope CTA will strongly improve indirect detection limits by 2030. Yet, astrophysical uncertainties constitutes a limitation. We can however hope for improvements in our knowledge of the halo profiles, local dark matter density and cosmic ray propagation models, which would lead to more robust constraints from direct and indirect detection experiments.